\documentclass[aps,prd,showpacs,twocolumn,nofootinbib,nosuperscriptaddress,amsmath,amssymb]{revtex4-2}

\usepackage{graphics}
\usepackage{epsfig}
\usepackage{graphicx}
\usepackage{color}
\usepackage{mathrsfs}
\usepackage{bm}
\usepackage{mathtools}
\usepackage{float}
\usepackage{amssymb}
\usepackage{braket}
\usepackage{fixltx2e}
\usepackage{bm}
\usepackage{amsmath,amsfonts,bm,dsfont}
\usepackage{verbatim}
\usepackage{mathrsfs}  %
\usepackage{definitions}

\newcommand{\rv}[1]{{#1}}
\newcommand{\zz}[1]{{#1}}
\def\rA{{\rm A}}
\def\rR{{\rm R}}
\def\rK{{\rm K}}
\def\rM{{\rm M}}
\def\rT{{\rm T}}

\begin{document}

\title{Precise Wigner-Weyl calculus for the honeycomb lattice}

\author{R. Chobanyan}
\affiliation{Physics Department, Ariel University, Ariel 40700, Israel}

\author{M.A. Zubkov \footnote{On leave of absence from Institute for Theoretical and Experimental Physics, B. Cheremushkinskaya 25, Moscow, 117259, Russia}}
\affiliation{Physics Department, Ariel University, Ariel 40700, Israel}

\date{\today}

\begin{abstract}
In this paper we propose the precise Wigner-Weyl calculus for the lattice models defined on the honeycomb lattice. We construct two symbols of operators: the $\mathscr{B}$-symbol, which is similar to the symbol introduced by F. Buot, and the $W$ (or, Weyl) symbol. The latter possesses the  set of useful properties. These identities allow us to use it in   physical applications. In particular, we derive topological expression for the Hall conductivity through the Wigner transformed Green function. This expression may be used for the description of quantum Hall effect in the systems with artificial honeycomb lattice, when magnetic flux through the lattice cell is of the order of elementary quantum of magnetic flux.   
\end{abstract}

\maketitle
\tableofcontents


\section{Introduction}
\label{SectIntro}

Wigner-Weyl calculus in its original form has been proposed  by  H. Groenewold \cite{Groenewold1946} and J. Moyal \cite{Moyal1949}. It was designed to replace operator formulation of quantum mechanics by phase space formulation, where the basic notion is Weyl symbol of operator, i.e. function in phase space that carries all necessary information about the operator itself. Here the ideas of H. Weyl \cite{Weyl1927} and E. Wigner \cite{Wigner1932} have been used, which gave to this calculus its present name.  Instead of the non - commutative operator product in the Wigner - Weyl formalism the non - commutative  Moyal product is used  \cite{Ali2005,Berezin1972}. The calculus had several applications to quantum mechanics \cite{Curtright2012,Zachos2005} and quantum field theory \cite{Cohen1966,Agarwal1970,E.C.1963,Glauber1963,Husimi1940,Cahill1969,Buot2009}.

Originally Wigner - Weyl formalism has been proposed for the systems defined in continuum space. It is well - known, however, that lattice regularization is necessary for the self - consistent definition of quantum field theory. Moreover, in solid state physics the tight - binding models give reasonable description of collective excitations. Therefore, the field theory systems defined on the lattice represent an important domain of condensed matter physics. The attempts to define Wigner-Weyl calculus for the lattice models have been undertaken long time ago, starting from the works of  Schwinger \cite{Schwinger570}. Important ideas in this direction have been proposed  by Buot \cite{Buot1974,Buot2009,Buot2013}. The corresponding constructions have been proposed by several authors, including Wooters \cite{WOOTTERS19871},  Leonhardt \cite{Leonhardt1995},   Kasperowitz \cite{KASPERKOVITZ199421}, and Ligab\'o \cite{Ligabo2016}, see also   \cite{BJORK2008469,GALETTI1988267,Cohendet_1988,doi,PhysRevA.53.3822,rivas1999weyl,mukunda2004wigner,chaturvedi2005wigner} and references therein. It is worth mentioning also that the large chapter of pure mathematics called deformational quantization is based on Wigner - Weyl calculus \cite{BAYEN197861,Kontsevich2003,Felder2000,Kupriyanov2008}.

With the purpose of applications to non - dissipative transport phenomena the so - called approximate lattice Wigner-Weyl calculus was suggested \cite{ZW2019}.
This version of Wigner - Weyl calculus may be applied to the lattice systems with  weak inhomogeneity caused by slowly varying external fields. In practise this formalism can be used for the consideration of any lattice regularized continuous field theory, and to the solid state systems in the presence of elastic deformations, weak disorder, and magnetic fields much smaller than $10^5$ Tesla. The latter requirement seems at a first look to be fulfilled always because the maximal values of magnetic fields accessed in laboratories do not exceed $100$ Tesla. With the aid of approximate Wigner - Weyl calculus the conductivities of several non  - dissipative transport effects have been expressed through the topological invariants  \cite{Zubkov2017,Chernodub2017,Khaidukov2017,Zubkov2018a,Zubkov2016a,Zubkov2016b,Chernodub2016,Chernodub2017}.
The consideration of essentially non - homogeneous systems within this methodology has been performed in  \cite{ZW2019,FZ2020,ZZ2021,BFLZZ2021}.

In the present paper we concentrate on the Quantum Hall Effect (QHE). Its topological description for the idealized system of non  - interacting electrons in the presence of constant external magnetic field has been proposed in \cite{Thouless1982}. The QHE conductivity in such systems is proportional to the so  - called TKNN invariant expressed through the integral of Berry curvature over the occupied energy levels. The TKNN invariant is robust to the smooth modification of the one  - particle Hamiltonian  \cite{Avron1983,Fradkin1991,Hatsugai1997,Qi2008,Kaufmann:2015lga,Tong:2016kpv}.
However, the  application of the TKNN invariant is limited to unphysically idealized systems without interactions and disorder.

 The alternative topological description of the QHE is given in terms of the Green functions. First the intrinsic anomalous QHE (existing without external magnetic field) in homogenous topological insulators without interactions  \cite{IshikawaMatsuyama1986,Volovik1988,Volovik2003a} has been given. The corresponding expression for the conductivity is given by
$$
\sigma_H = \frac{e^2}{h}{\cal N},
$$
where
\be
{\cal N}
=  -\frac{ \epsilon_{ijk}}{  \,3!\,4\pi^2}\, \int d^3p \Tr
\[
{G}(p ) \frac{\partial {G}^{-1}(p )}{\partial p_i}  \frac{\partial  {G}(p )}{\partial p_j}  \frac{\partial  {G}^{-1}(p )}{\partial p_k}
\].
\label{N-0}
\ee
In this expression $G(p)$ is the two - point Green function in momentum space.
The advantage of Eq. (\ref{N-0}) is that contrary to the TKNN invariant it may be extended to the systems with interactions. Then the two-point Green function is taken with the interaction corrections \cite{ColemanHill1985,Lee1986,ZZ2019}. Notice that the role of  interaction corrections to the QHE conductivity was considered long time ago, well before the mentioned description with the aid of Eq. (\ref{N-0})    \cite{KuboHasegawa1959,Niu1985a,Altshuler0,Altshuler}.

Eq. (\ref{N-0}) solves the problem with interaction corrections to the QHE in topological insulators. However, it, strictly speaking, cannot be used for the description of ordinary QHE in the presence of external magnetic field, and, more widely, for the QHE in the non - homogeneous systems. The extension of this expression to the non - homogeneous systems has been given in \cite{ZW2019}. For the tight - binding model of a two - dimensional lattice system
the Hall conductivity averaged over the system area is   $
\sigma_H = \frac{\cal N}{2\pi},
$ with
\begin{widetext}
\be
{\cal N}
= - \frac{T \epsilon_{ijk}}{ |{\bf A}| \,3!\,4\pi^2}\, \int \D{{}^3x} \int_{\cM}  \D{{}^3p}
\, {\rm tr}\, {G}_{\cC}(x,p )\star \frac{\partial {Q}_{\cC}(x,p )}{\partial p_i} \star \frac{\partial  {G}_{\cC}(x,p )}{\partial p_j} \star \frac{\partial  {Q}_{\cC}(x,p )}{\partial p_k}
\label{calM2d230I}
\ee
\end{widetext}
Here $T \to 0$ is temperature, $|{\bf A}| \to \infty $ is the total area of the system, ${G}_{\cC}(x,p )$ is Wigner transformation of the two-point Green function $\hat G$. ${Q}_{\cC}(x,p )$ is lattice Weyl symbol of operator $\hat{Q}$. Here $\hat{Q}$ is operator inverse to the Green function. Moyal product is denoted by $\star$. As it was mentioned above, here the approximate version of lattice Wigner - Weyl calculus has been used, which allows to deal with realistic magnetic fields much smaller than $10^5$ Tesla.
It is worth mentioning that by definition
 the Weyl symbol is defined as a function of real valued coordinates, not only for the   discrete lattice points.
In \cite{ZZ2019_2,ZZ2021} it was proven that Eq. (\ref{calM2d230I}) remains valid in the presence of interactions if bare non - interacting Green function is replaced by the complete interacting one.

Eq. (\ref{calM2d230I}) is to be modified for the artificial lattices or in case of strong inhomogeneities. These are the systems, where the so - called Hofstadter butterfly appears. Extension of  Eq. (\ref{calM2d230I}) to such systems has been given in \cite{FZ2020}, where the consideration was limited by the tight - binding models defined on the infinite rectangular lattice. The corresponding version of lattice Wigner - Weyl calculus  was called "precise" because the corresponding Weyl symbol of an operator satisfies the basic identities of continuous Wigner - Weyl calculus precisely. Within this formalism the expression for the QHE conductivity has been derived, where Eq. (\ref{calM2d230I}) is  replaced by
\begin{widetext}
\be
{\cal N}
= - \frac{ \epsilon_{ijk}}{ |{\bf A}| \,3!\,4\pi^2}\, \frac{|{\cal V}^{(2)}|}{2^{2} }\sum_{\overrightarrow x \in {\mathscr{D}}} \int_{\cM} {d^3p} \Tr
\[
{G}_{W}(x,p )\star \frac{\partial {Q}_{W}(x,p )}{\partial p_i} \star \frac{\partial  {G}_{W}(x,p )}{\partial p_j} \star \frac{\partial  {Q}_{W}(x,p )}{\partial p_k}
\].
\label{calM2d230}
\ee
\end{widetext}
Here $x = (\tau, \overrightarrow{x})$, $\overrightarrow{x}$ is the point in space, $\tau$ is imaginary time that belongs to the interval between $0$ and $1/T \to \infty$. Wigner transformation of Green function ${G}_{W}$ and Weyl symbol of Dirac operator ${Q}_{W}$ do not depend on $\tau$.  ${\cal V}^{(2)}$ is the area of the elementary lattice cell, $|{\bf A}|\to \infty$ is the system area. By $\mathscr{D}$ we denote the extended  lattice, in which the extra lattice sites are added with the half - integer coordinates (assuming the coordinates of the original lattice are integer).

Recall that Eq. (\ref{calM2d230}) has been derived specifically for the systems defined on the rectangular lattice. In practise, however, the systems with large magnetic flux through the lattice cell have been obtained with the artificial lattices that do not have the rectangular form (see, for example, \cite{dean2013hofstadter} and references therein). In particular, the artificial lattices may have the honeycomb form \cite{polini2013artificial}. Therefore, in the present paper we extend the precise Wigner - Weyl calculus to the systems defined on the honeycomb lattices. We found that it is possible to overcome technical difficulties specific for the honeycomb lattice, and we arrive at the Wigner - Weyl calculus with the definition of the Weyl symbol  that obeys basic properties of the continuous Wigner - Weyl calculus:

\begin{itemize}
    \item Star product identity
    \begin{equation}
        (AB)_{W}(x,p) = A_{W}(x,p) \star B_{W}(x,p)
        \label{01}
    \end{equation}
    \item First trace identity
    \begin{equation}
        {\rm Tr}A_{W} = {\rm tr}\hat{A}
        \label{02}
    \end{equation}
    \item Second trace identity
    \begin{equation}
        {\rm Tr}\left[ A_{W}(x,p)B_{W}(x,p) \right] = {\rm Tr}\left[ A_{W}(x,p) \star B_{W}(x,p) \right]
        \label{03}
    \end{equation}
    \item Weyl symbol of the identity operator
    \begin{equation}
        \left( \hat{1} \right)_{W}(x,p) = 1
        \label{04}
    \end{equation}
\end{itemize}
where the Moyal product, as defined below, serves as the star product mentioned in property (\ref{01}):
    \begin{equation}
        \star \equiv \star_{x,p} = e^{\frac{i}{2} \left( \overleftarrow{\partial_{x}}\overrightarrow{\partial_{p}} - \overleftarrow{\partial_{p}}\overrightarrow{\partial_{x}} \right)}
        \label{05}
    \end{equation}

Using the designed precise Wigner - Weyl calculus we derive expression for the Hall conductivity of the system defined on the honeycomb lattice. We rely on the unification of the lattice Wigner - Weyl calculus with Keldysh technique. Here we follow the methodology developed in \cite{Sugimoto,Sugimoto2006,Sugimoto2007,Sugimoto2008,BFLZZ2021}.


\section{Statement of the main results}
\label{Statement}
\subsection{Definition of Weyl symbol and its properties}
\label{SectList}

We are considering the honeycomb lattice of monolayer graphene. The physical lattice vectors of graphene are defined as follows:
\begin{gather}
	\Vec{l}_{1} = \frac{\ell}{2} \left( 3\hat{x} + \sqrt{3}\hat{y} \right) \nonumber \\
	\Vec{l}_{2} = \frac{\ell}{2} \left( 3\hat{x} - \sqrt{3}\hat{y} \right) \nonumber \\
	\Vec{\ell} = \ell\hat{x}
	\label{G01}
\end{gather}
where the last vector is the representation of the basis. These are the reciprocal lattice vectors:
\begin{gather}
	\Vec{g}_{1} = \frac{2\pi}{3\ell} \left( \hat{k}_{x} + \sqrt{3}\hat{k}_{y} \right) \nonumber \\
	\Vec{g}_{2} = \frac{2\pi}{3\ell} \left( \hat{k}_{x} - \sqrt{3}\hat{k}_{y} \right)
	\label{G02}
\end{gather}

The physical lattice is defined by the following:
\begin{equation}
	\mathscr{O} \equiv
	\begin{rcases}
		\begin{dcases}
			2c_{1}^{1}\Vec{l}_{1} + 2c_{2}^{1}\Vec{l}_{2}, & c_{1,2}^{1} \in{\mathbb{Z}} \\
			2\Vec{\ell} + 2c_{1}^{2}\Vec{l}_{1} + 2c_{2}^{2}\Vec{l}_{2}, & c_{1,2}^{2} \in{\mathbb{Z}}
		\end{dcases}
	\end{rcases}
	\label{GA01}
\end{equation}
The physical lattice's first Brillouin zone is given by:
\begin{equation}
	\mathscr{M} = \left\{ \frac{1}{2}m_{1} \Vec{f}_{1} + \frac{1}{2}m_{2} \Vec{f}_{2},\;
	\begin{dcases}
		m_{1} \in \left( -1/2,1/2 \right] \\
		m_{2} \in \left( -1/4,1/4 \right]
	\end{dcases} \right\}
	\label{GA02}
\end{equation}

The extended lattice, which is the union of the physical and auxiliary lattices, is:
\begin{equation}
	\mathfrak{D} \equiv
	\begin{rcases}
		\begin{dcases}
			c_{1}^{1}\Vec{l}_{1} + c_{2}^{1}\Vec{l}_{2}, & c_{1,2}^{1} \in{\mathbb{Z}} \\
			-\Vec{\ell} + c_{1}^{2}\Vec{l}_{1} + c_{2}^{2}\Vec{l}_{2}, & c_{1,2}^{2} \in{\mathbb{Z}}
		\end{dcases}
	\end{rcases} = \mathscr{O}\cup\mathscr{O'}
	\label{GA04}
\end{equation}
The extended lattice's first Brillouin zone is given by:
\begin{equation}
	\mathfrak{M} = \left\{ m_{1} \Vec{f}_{1} + m_{2} \Vec{f}_{2},\;
	\begin{dcases}
		m_{1} \in \left( -1/2,1/2 \right] \\
		m_{2} \in \left( -1/4,1/4 \right]
	\end{dcases} \right\}
	\label{GA05}
\end{equation}


For $x \in \mathfrak{D}, p \in \mathscr{M}$ we define Weyl symbol of operator $\hat A$ as follows
 \begin{align}
	A_{W}(x,p) &\equiv  \int_{\mathscr{M}} d^{2}q e^{2ixq} \langle p + q|\hat{A}|p - q\rangle \nonumber \\
	&\times \left( 1 + e^{-2il_{1}q} + e^{-2il_{2}q} + e^{-2i(l_{1} + l_{2})q} \right)
	\label{GG01}
\end{align}
This expression is naturally extended to continuous values of $x$.

We formulate the following properties of the Weyl symbol:

\begin{enumerate}
	\item
	
	Trace property.
	$$	
	{\rm Tr}\, \hat{A} = \sum_{x \in \mathfrak{D}} \int_{\mathscr{M}} \frac{d^{2}p}{|\mathfrak{M}|} A_W(x,p)
	$$	
	
	\item

	Second trace identity.
	
	$$	
	{\rm Tr} \hat{A}\hat{B} = \sum_{x \in \mathfrak{D}} \int_{\mathscr{M}} \frac{d^{2}p}{|\mathfrak{M}|} A_W(x,p) B_W(x,p)
	$$

	\item Star property
	
	\begin{eqnarray}
		&&	(\hat{A}\hat{B})_W(x,p)\Big|_{p \in {\cal M}, x\in \mathfrak{D} }
	\nonumber\\&&	=  A_W(p,q)  e^{\frac{i}{2}(\overleftarrow{\partial_q}\overrightarrow{\partial_p}-\overleftarrow{\partial_p}\overrightarrow{\partial_q})}  B_W(p,q) \label{star0}
	\end{eqnarray}
	
	\item Weyl symbol of unity
	
	$$
	1_W(x,p)\Big|_{p \in {\cal M} , x\in \mathfrak{D} }      = 1
	$$
	
	\item Star product without differentiation
	
	\begin{eqnarray}
		&&A_{W}(x,p) \star B_{W}(x,p)\Big|_{x \in \mathscr{D}}
		=
		\sum_{z,\bar{z} \in \mathscr{D}} \int \frac{\D{p}^\prime}{|{\mathfrak{M}}|}\frac{d\bar{p}^\prime}{|{\mathfrak{M}}|}\nonumber\\&&e^{ 2ip^\prime (\bar{z}-x) + 2i \bar{p}^\prime (x-z)} A_{W}(z,p-p^\prime)B_{W}(\bar{z},p-\bar{p}^\prime)
	\end{eqnarray}

\end{enumerate}

\subsection{Quantum Hall effect in  condensed matter system defined on honeycomb lattice}

We consider inhomogeneous system defined on the honeycomb lattice $\mathscr{O}$. Time remains continuous. The fermionic field $\Phi$ is defined on lattice sites. We assume that the system remains non - interacting. Keldysh Green function is defined as the following matrix
\begin{eqnarray}
&&	\hat{ G}(t,x|t^\prime,x^\prime)
\nonumber\\&&	= -i \left(\begin{array}{cc}\langle T \Phi(t,x) \Phi^+(t^\prime,x^\prime)\rangle & -\langle  \Phi^+(t^\prime,x^\prime) \Phi(t,x)\rangle\\ \langle  \Phi(t,x) \Phi^+(t^\prime,x^\prime)\rangle & \langle \tilde{T} \Phi(t,x) \Phi^+(t^\prime,x^\prime)\rangle \end{array} \right).
	\label{KelG_S}
\end{eqnarray}
Here Heisenberg fermionic field operator $\Phi$ depends on time $t$ and spatial coordinates $x$. By $T$ we denote the time ordering while $\tilde{T}$ is anti - time ordering.

The $2+1$ dimensional vectors (with space and time components) are denoted by large Latin letters.  Correspondingly, in momentum space $A(P_1,P_2) = \langle P_1 | \hat{A} | P_2 \rangle $. The space components of momentum belong to the Brillouin zone while its time component (frequency) is real - valued. We then define Weyl symbol of an operator $\hat A$ as the mixture of lattice Weyl symbol and Wigner transformation with respect to the frequency component:
\begin{widetext}
\begin{eqnarray}
	A_W(X|P)&=&2\int d P^0\,\int_{\mathscr{M}} d^{2}\vec{Q}\,  e^{-2 \ii X^\mu Q_\mu }  A(P+Q,P-Q) \left( 1 + e^{-2il_{1}\vec{Q}} + e^{-2il_{2}\vec{Q}} + e^{-2i(l_{1} + l_{2})\vec{Q}} \right) \label{WignerTr} \\ && \quad \mu =0,1,2\nonumber
\end{eqnarray}
\end{widetext}
$2+1$ momentum is denoted by $P^\mu=(P^0,p)$, and $P_\mu = (P^0,-p)$. Here  $p$ is spatial momentum with $D$ components.
Keldysh Green function is an operator inverse to $\hat{\bf Q}$.
Weyl symbol of Keldysh Green function $\hat{\bf G}$ is denoted by $\hat{G}_W$, while Weyl symbol of Keldysh $\hat{\bf Q}$ is $\hat{Q}_W$.

We obtain the following results for the dynamics of lattice system written in terms of Weyl symbols of operators.

\begin{enumerate}
	\item

	Weyl symbols $\hat{G}_W$ and $\hat{Q}_W$ obey Groenewold equation
	\begin{equation}
		\hat{Q}_W * \hat{G}_W = 1_W.
	\end{equation}
	Here the Moyal product $*$ is defined as
	\begin{eqnarray}
		&&\left(A* B\right)(X|P)\nonumber\\&& = A(X|P)\,e^{\rv{-}\ii(\overleftarrow{\partial}_{X^{\mu}}\overrightarrow{\partial}_{P_{\mu}}-\overleftarrow{\partial}_{P_{\mu}}\overrightarrow{\partial}_{X^{\mu}})/2}B(X|P).
	\end{eqnarray}
	
	It is worth mentioning that for the complete description of the system we need the values of Weyl symbols defined on  spatial phase space $\mathfrak{D}\otimes \mathscr{M}$. For such values of spatial momenta and coordinates the Weyl symbol of unity is equal to $1$.
	
	\item

	We express DC conductivity (in units of $e^2/\hbar$, averaged over the lattice area) of the two - dimensional {\it non - interacting} systems as
\begin{widetext}	
	\begin{equation}
		\sigma^{ij} =  {\frac{1}{4}} \frac{1}{|\mathfrak{D}|} \int \frac{dP^0}{2\pi} \int_{\mathscr{M}} \frac{d^{2}\vec{P}}{(2\pi)^2}\, \sum_{x\in \mathfrak{D}} \tr\left(\partial_{\pi_{i}}\hat{Q}_W  \left[\hat{G}_W \star \partial_{\rv{\pi_{[0}}}\hat{Q}_W  \star \partial_{\rv{\pi_{j]}}}\hat{G}_W  \right]\right)^< +{\rm c.c.}\label{MAIN}
	\end{equation}
\end{widetext}
	through the lesser component of expression that contains Wigner transformed Keldysh Green function $G(X|\pi)$ and its inverse $Q$ that obey Groenewold equation $Q \star G = 1$. Here by $|\mathfrak{D}|$ we denote the number of lattice points in the extended lattice $\mathfrak{D}$. It is assumed here that this number is large but remains finite. Using representation of Keldysh Green function of Eq. (\ref{KelG_S}) the lesser component in the above representation may be rewritten explicitly as
	\begin{widetext}
	\begin{equation}
		\sigma^{ij} =   {\frac{1}{4}} \frac{1}{|\mathfrak{D}|} \int \frac{dP^0}{2\pi} \int_{\mathscr{M}} \frac{d^{2}\vec{P}}{(2\pi)^2}\, \sum_{x\in \mathfrak{D}} \tr\left( \gamma^<\partial_{\pi_{i}}\hat{Q}_W  \left[\hat{G}_W \star \partial_{\rv{\pi_{[0}}}\hat{Q}_W  \star \partial_{\rv{\pi_{j]}}}\hat{G}_W  \right]\right) +{\rm c.c.}\label{MAIN}
	\end{equation}
\end{widetext}
	where trace is taken over Keldysh components as well as over the internal indices, while
	\zz{$$
		\gamma^< = \left(\begin{array}{cc} 0 & 0 \\ 1 & 0 \end{array} \right).
		$$}
	Here $(...)_{[0} (...)_{ j]} =(...)_{0} (...)_{ j} -(...)_{j} (...)_{ 0}  $ means anti-symmetrization.

	\item

	We show that the above expression for the conductivity (in units of $e^2/\hbar$, averaged over the system area) is reduced to the following expression in case of the equilibrium system at zero temperature
	$$
	\bar{\sigma}^{ij}  = \frac{{\cal N}}{2\pi}\epsilon^{ij},
	$$
	where
	\begin{widetext}
	\begin{eqnarray}
		{\cal N} &=& \frac{1}{24\,\pi^2}\epsilon^{\mu\nu\rho} \frac{1}{|\mathfrak{D}|} \int {dP^0} \int_{\mathscr{M}} {d^{2}\vec{P}}\, \sum_{x\in \mathfrak{D}}
		\tr\left(\partial_{\rv{\Pi^{\mu}}}\hat{Q}_W ^\rM  \star\hat{G}_W ^\rM \star \partial_{\rv{\Pi^{\nu}}}\hat{Q}_W ^\rM  \star\hat{G}_W ^\rM \star \partial_{\rv{\Pi^{\rho}}}\hat{Q}_W ^\rM \star \hat{G}_W ^\rM \right).
		\label{NEQ0}
	\end{eqnarray}
\end{widetext}
	Here the $\hat{G}_W^M$ is the  Weyl symbol of  the Matsubara Green function while $\hat{Q}_W^M$ is Weyl symbol of its inverse.
	Momentum space is Euclidean one, its points are denoted by $\Pi_i$. $\Pi^3$ is Matsubara frequency.

	\item

	One can check that Eq. (\ref{NEQ0}) is a topological invariant. For that we need that the system was in thermal equilibrium originally, and that the thermal equilibrium corresponds to zero temperature. Moreover, we need that the Hamiltonian does not depend on time. The value of the average conductivity $\bar{\sigma}^{ij}$ is then robust to smooth variations of the system. The sum over $x$ is important for the topological invariance of this quantity.

\end{enumerate}

\section{Wigner-Weyl calculus of Felix Buot (1D)}
\label{Sect1DBuot}

The precise Wigner-Weyl calculus for lattice models inspired by the original construction of F. Buot is excerpted here. Notice that strictly speaking, the Buot symbol of an operator differs from the original definition given by F. Buot. We call it the Buot symbol to distinguish it from the construction that uses the extended lattice. We restrict ourselves to a simple one-dimensional physical lattice $\mathscr{O}$ with a one-dimensional first Brillouin zone $\mathscr{M}$.

\subsection{The Hilbert space}
\label{Sect01}

The following definition applies to the physical lattice, which is represented by the symbol $\mathscr{O}$:
    \begin{equation}
        \mathscr{O} \equiv \{2\ell k,\; k\in{\mathbb{Z}}\}
        \label{A01}
    \end{equation}
where $2\ell$ stands for the lattice spacing. Its first Brillouin zone is denoted by the symbol $\mathscr{M}$ and has the following definition:
    \begin{equation}
        \mathscr{M} = \left( -\frac{\pi}{2\ell},\frac{\pi}{2\ell} \right]
        \label{A02}
    \end{equation}

Also, the auxiliary lattice, which is a translation of the physical lattice by half the lattice spacing, is defined as follows:
    \begin{equation}
        \mathscr{O'} \equiv \{\ell(2k + 1),k\in{\mathbb{Z}}\} = \mathscr{O} + \ell
        \label{A03}
    \end{equation}

Consequently, it is possible to define the extended lattice, represented by the symbol $\mathfrak{D}$, which comprises both the auxiliary lattice and the physical lattice, as follows:
    \begin{equation}
        \mathfrak{D} \equiv \{\ell k,\; k\in \mathbb{Z}\} = \mathscr{O} \cup \mathscr{O'}
        \label{A04}
    \end{equation}
Here, $\ell$ stands for the lattice spacing, and $\mathfrak{M}$ is the definition of the first Brillouin zone of the extended lattice:
    \begin{equation}
        \mathfrak{M} = \left( -\frac{\pi}{\ell},\frac{\pi}{\ell} \right]
        \label{A05}
    \end{equation}

The characteristics of the physical states are described by the following definitions:
    \begin{gather}
        \hat{1}_{\mathscr{O}} = \sum_{x \in \mathscr{O}} |x\rangle \langle x| = \int_{\mathscr{M}} dp |p\rangle \langle p| \qquad \langle x|p \rangle = \frac{1}{\sqrt{|\mathscr{M}|}}e^{ixp} \nonumber \\
        \langle p|q \rangle = \delta^{\left[ \frac{\pi}{\ell} \right]}(p - q) \qquad \langle x|y \rangle = \delta_{x,y} \nonumber \\
        |p \rangle = \frac{1}{\sqrt{|\mathscr{M}|}}\sum_{x\in\mathscr{O}} e^{ixp}|x\rangle
        \label{A06}
    \end{gather}
The latter being the Fourier transform.


\subsection{The $\mathscr{B}$-symbol}
\label{Sect2}

The Wigner transformation of a function $B(p,q)$, where $p,q \in \mathscr{M}$, is defined as follows:
    \begin{align}
        B_{\mathscr{B}}(x,p) & \equiv \frac{1}{2} \int_{\mathfrak{M}} dq e^{ixq}B \left( p + \frac{q}{2},p - \frac{q}{2} \right) \nonumber \\
        &= \int_{\mathscr{M}} dq e^{2ixq}B(p + q,p - q)
        \label{B01}
    \end{align}

In addition, the inverse transformation of a function $Q_{\mathscr{B}}(x,p)$ is given by:
    \begin{equation}
        Q(p,q) = \frac{1}{|\mathscr{M}|}\sum_{x \in \mathfrak{D}} e^{-i(p - q)x}Q_{\mathscr{B}} \left( x,\frac{p + q}{2} \right)
        \label{B02}
    \end{equation}
Using the Fourier transform in (\ref{A06}) and the $\mathscr{B}$-symbol definition in (\ref{B01}), the following can be demonstrated:
    \begin{align*}
        &\frac{1}{|\mathscr{M}|} \sum_{x \in \mathfrak{D}} e^{-ikx}Q_{\mathscr{B}}(x,p) = \frac{1}{|\mathscr{M}|} \sum_{x \in \mathfrak{D}} e^{-ikx} \int_{\mathscr{M}} dq e^{2ixq} \nonumber \\
        &\times Q(p + q,p - q) \nonumber \\
        &= \frac{1}{|\mathscr{M}|} \sum_{x \in \mathfrak{D}} e^{2ix \left(q - \frac{k}{2} \right)} \int_{\mathscr{M}} Q(p + q,p - q) dq \nonumber \\
        &= \frac{1}{|\mathscr{M}|} \sum_{n \in \mathbb{Z}} e^{in\ell (2q - k)} \int_{\mathscr{M}} Q(p + q,p - q) dq \nonumber \\
        &= \frac{2\pi}{|\mathscr{M}|} \sum_{n \in \mathbb{Z}} \delta \left[ \ell(2q - k) -2\pi n \right] \int_{\mathscr{M}} Q(p + q,p - q) dq \nonumber \\
        &= \frac{2\pi}{|\mathscr{M}|} \frac{1}{2\ell} \sum_{n \in \mathbb{Z}} \delta \left( q - \frac{k}{2} -\frac{\pi}{\ell}n \right) \int_{\mathscr{M}} Q(p + q,p - q) dq   \nonumber \\
        &= \int_{\mathscr{M}} dq \delta^{\left[ \frac{\pi}{\ell} \right]} \left( q - \frac{k}{2} \right) Q(p + q,p - q) \nonumber \\
        &= Q \left(p + \frac{k}{2},p - \frac{k}{2} \right) \tag*{$\blacksquare$}
    \end{align*}

\subsection{Moyal product}
\label{Sect3}

\begin{figure}[H]
    \centering
        \begin{minipage}{0.45\textwidth}
            \centering
            \includegraphics[scale = 0.5]{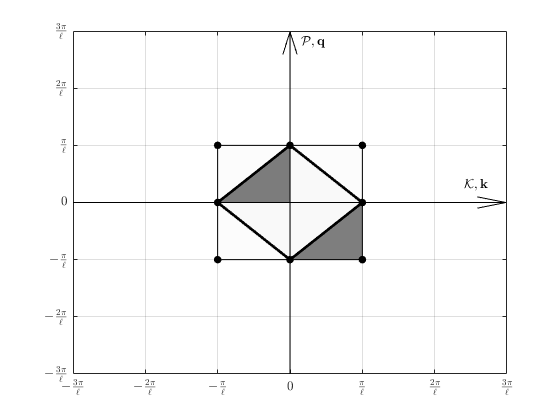}
            \caption{The integral over a square is transformed into a rhombus; the contributions from shaded areas are equal.}
            \label{fig:star}
        \end{minipage}
\end{figure}

It is possible to show that the $\mathscr{B}$-symbol of an operator obeys the star-product identity stated in Sect. \ref{SectIntro}-(\ref{01}) through the definition of the $\mathscr{B}$-symbol in (\ref{B01}):
    \begin{align*}
        &(AB)_{\mathscr{B}}(x,p) = \frac{1}{2} \int_{\mathfrak{M}} d\mathcal{P} e^{ix\mathcal{P}} \left< p + \frac{\mathcal{P}}{2} \left| \hat{A}\hat{B} \right| p - \frac{\mathcal{P}}{2} \right> \nonumber \\
        &= \frac{1}{2} \int_{\mathfrak{M}} d\mathcal{P} e^{ix\mathcal{P}} \int_{\mathscr{M}} d\mathcal{R} \left< p + \frac{\mathcal{P}}{2} \left| \hat{A} \right| \mathcal{R} \right> \left< \mathcal{R} \left| \hat{B} \right| p - \frac{\mathcal{P}}{2} \right> \nonumber \\
        &= \frac{1}{4} \int_{\mathfrak{M}} d\mathcal{P} d\mathcal{K} e^{ix\mathcal{P}} \left< p + \frac{\mathcal{P}}{2} \left| \hat{A} \right| p - \frac{\mathcal{K}}{2} \right> \nonumber \\
        &\times \left< p - \frac{\mathcal{K}}{2} \left| \hat{B} \right| p - \frac{\mathcal{P}}{2} \right> \nonumber \\
        &= \frac{2}{2 \times 4} \int_{\mathfrak{M}} dq dk e^{ix(q + k)} \left< p + \frac{q}{2} + \frac{k}{2} \left| \hat{A} \right| p - \frac{q}{2} + \frac{k}{2} \right> \nonumber \\
        &\times \left< p - \frac{q}{2} + \frac{k}{2} \left| \hat{B} \right| p - \frac{q}{2} - \frac{k}{2} \right> \nonumber \\
        &= \frac{1}{4} \int_{\mathfrak{M}} dq dk e^{ixq} \left< p + \frac{q}{2} \left| \hat{A} \right| p - \frac{q}{2} \right> e^{\frac{k}{2} \overleftarrow{\partial_{p}} - \frac{q}{2}\overrightarrow{\partial_{p}}} \nonumber \\
        &\times e^{ixk} \left< p + \frac{k}{2} \left| \hat{B} \right| p - \frac{k}{2} \right> \nonumber \\
        &= \frac{1}{2} \int_{\mathfrak{M}} dq e^{ixq} \left< p + \frac{q}{2} \left| \hat{A} \right| p - \frac{q}{2} \right> e^{\frac{i}{2} \left( \overleftarrow{\partial_{x}}\overrightarrow{\partial_{p}} - \overleftarrow{\partial_{p}}\overrightarrow{\partial_{x}} \right)} \nonumber \\
        &\times \frac{1}{2} \int_{\mathfrak{M}} dk e^{ixk} \left< p + \frac{k}{2} \left| \hat{B} \right| p - \frac{k}{2} \right> \tag*{$\blacksquare$}
    \end{align*}
The factor $2$ in the fifth line is the Jacobian generated by changing the variables $\mathcal{P} = q + k$ and $\mathcal{K} = q - k$, where $q,k\in \mathfrak{M}$. Additionally, changing the variables transforms the integration region contained within $\mathfrak{M} \times \mathfrak{M}$ from a square to a rhomboid, as shown in FIG. \ref{fig:star}. Due to the periodicity of the integral over the rhomboid form, the factor $1/2$ appears in the fifth line when the integration region is transformed back from a rhomboid to a square.

\subsection{Trace and its properties}
\label{Sect4}

The definitions of the traces of the physical and extended lattices are as follows:
    \begin{align}
        &{\rm Tr}_{\mathscr{O}} A_{\mathscr{B}} \equiv \sum_{x \in \mathscr{O}} \int_{\mathscr{M}} \frac{dp}{|\mathscr{M}|} A_{\mathscr{B}}(x,p) \nonumber \\
        &{\rm Tr}_{\mathfrak{D}} A_{\mathscr{B}} \equiv \sum_{x \in \mathfrak{D}} \int_{\mathscr{M}} \frac{dp}{|\mathscr{M}|} A_{\mathscr{B}}(x,p)
        \label{D01}
    \end{align}
Both adhere to the first trace identity stated in Sect. \ref{SectIntro}-(\ref{02}). As proof, consider the following:
    \begin{align*}
        &{\rm Tr}_{\mathfrak{D}}A_{\mathscr{B}} \equiv \frac{1}{|\mathscr{M}|}\sum_{x\in \mathfrak{D}} \int_{\mathscr{M}} dp A_{\mathscr{B}}(x,p) \nonumber \\
        &= \frac{1}{|\mathscr{M}|}\sum_{x\in \mathfrak{D}} \int_{\mathscr{M}} dp dqe^{2ixq} \langle p + q|\hat{A}|p - q \rangle \nonumber \\
        &= \frac{1}{|\mathscr{M}|}\sum_{x\in \mathfrak{D}} \sum_{x,y\in \mathscr{O}} \int_{\mathscr{M}} dp dqe^{2ixq} \langle p + q|z \rangle \langle z|\hat{A}|y \rangle \langle y|p - q \rangle \nonumber \\
        &= \frac{1}{|\mathscr{M}|^{2}}\sum_{x\in \mathfrak{D}} \sum_{z,y\in \mathscr{O}} \int_{\mathscr{M}} dpdq e^{i(2x - z - y)q} e^{i(y - z)p} \langle z|\hat{A}|y \rangle \nonumber \\
        &= \sum_{x\in \mathfrak{D}} \sum_{y,z\in \mathscr{O}} \delta_{2x,y + z}\delta_{y,z} \langle z|\hat{A}|y \rangle \nonumber \\
        &= \sum_{x\in \mathscr{O}} \langle x|\hat{A}|x \rangle = {\rm tr}\hat{A} \tag*{$\blacksquare$}
    \end{align*}
The same can be demonstrated for the physical lattice, but only the extended lattice accommodates the second trace identity specified in Sect. \ref{SectIntro}-(\ref{03}). The proof:
    \begin{align*}
        &{\rm Tr}_{\mathfrak{D}}(A_{\mathscr{B}} \star B_{\mathscr{B}}) \equiv \sum_{x \in \mathfrak{D}} \int_{\mathscr{M}} \frac{dp}{|\mathscr{M}|} A_{\mathscr{B}}(x,p) \star B_{\mathscr{B}}(x,p) \nonumber \\
        &= \frac{1}{|\mathscr{M}|} \sum_{x \in \mathfrak{D}} \int_{\mathscr{M}} dp dq dk e^{2ixq} \langle p + q|\hat{A}|p - q \rangle \nonumber \\
        &\times e^{\frac{i}{2} \left( \overleftarrow{\partial_{x}}\overrightarrow{\partial_{p}} - \overleftarrow{\partial_{p}}\overrightarrow{\partial_{x}} \right)} e^{2ixk} \langle p + k|\hat{B}|p - k \rangle \nonumber \\
        &= \frac{1}{|\mathscr{M}|^{3}} \sum_{x \in \mathfrak{D}} \sum_{y,z,\Bar{y},\Bar{z}\in \mathscr{O}} \int_{\mathscr{M}} dp dq dk e^{i(2x - y - z)q + i(z- y)p} \nonumber \\
        &\times \langle y|\hat{A}|z \rangle e^{\frac{i}{2} \left( \overleftarrow{\partial_{x}}\overrightarrow{\partial_{p}} - \overleftarrow{\partial_{p}}\overrightarrow{\partial_{x}} \right)} e^{i(2x - \Bar{y} - \Bar{z})k + i(\Bar{z} - \Bar{y})p} \langle \Bar{y}|\hat{B}|\Bar{z} \rangle \nonumber \\
        &= \frac{1}{|\mathscr{M}|^{3}} \sum_{x \in \mathfrak{D}} \sum_{y,z,\Bar{y},\Bar{z}\in \mathscr{O}} \int_{\mathscr{M}} dp dq dk e^{i(2x - y - z)q + i(z- y)p} \nonumber \\
        &\times \langle y|\hat{A}|z \rangle e^{i(\Bar{y} - \Bar{z})q + i(z - y)k} e^{i(2x - \Bar{y} - \Bar{z})k + i(\Bar{z} - \Bar{y})p} \langle \Bar{y}|\hat{B}|\Bar{z} \rangle \nonumber \\
        &= \sum_{x \in \mathfrak{D}} \sum_{y,z,\Bar{y},\Bar{z} \in \mathscr{O}} \delta_{2x + \Bar{y} - \Bar{z},y + z} \langle y|\hat{A}|z \rangle \delta_{y - z,\Bar{z} - \Bar{y}} \nonumber \\
        &\times \delta_{2x - y + z,\Bar{y} + \Bar{z}} \langle \Bar{y}|\hat{B}|\Bar{z} \rangle \nonumber \\
        &= \sum_{x\in \mathscr{O}} \langle x|\hat{A}\hat{B}|x \rangle = {\rm tr}\hat{A}\hat{B} \tag*{$\blacksquare$}
    \end{align*}

\subsection{The $\mathscr{B}$-symbol of the identity operator}
\label{Sect5}

The identity operator's $\mathscr{B}$-symbol is given by:
    \begin{equation}
        (\hat{1})_{\mathscr{B}} = \frac{1}{2} \left[ 1 + \cos{\frac{\pi}{\ell}x} \right]
        \label{E01}
    \end{equation}
where $x\in \mathfrak{D}$. Contrary to how it is shown in Sect \ref{SectIntro}-(\ref{04}), the identity operator's $\mathscr{B}$-symbol fluctuates and isn't necessarily unitary. The following provides proof for this:
    \begin{align*}
        &(\hat{1})_{\mathscr{B}} = \int_{\mathscr{M}} dq e^{2ixq}\langle p + q|p - q\rangle \nonumber \\
        &= \int_{-\frac{\pi}{2\ell}}^{\frac{\pi}{2\ell}} dq e^{2ixq}\delta^{\left[ \frac{\pi}{\ell} \right]}(2q) = \frac{1}{2} \int_{-\frac{\pi}{2\ell}}^{\frac{\pi}{2\ell}} dq e^{2ixq}\delta^{\left[ \frac{\pi}{2\ell} \right]}(q) \nonumber \\
        &= \frac{1}{2} \int_{-\frac{\pi}{2\ell}}^{\frac{\pi}{2\ell}} dq e^{2ixq} \left[ \delta(q) + \frac{1}{2}\delta \left( q - \frac{\pi}{2\ell}\right) + \frac{1}{2}\delta \left( q + \frac{\pi}{2\ell} \right) \right] \nonumber \\
        &= \frac{1}{2} + \frac{1}{4}e^{i\frac{\pi}{\ell}x} + \frac{1}{4}e^{-i\frac{\pi}{\ell}x} = \frac{1}{2} + \frac{1}{2}\cos{\frac{\pi}{\ell}x} \tag*{$\blacksquare$}
    \end{align*}
Because only half of each function exists inside the integration region $\mathscr{M}$, the shifted Dirac delta functions, enclosed in square parentheses in the third line, are factored by $1/2$.



\section{Wigner-Weyl calculus of Felix Buot in graphene}
\label{SectGraphBuot}

Carbon atoms are arranged in a honeycomb lattice to form a monolayer called graphene. It is possible to think of this physical lattice as the union of two Bravais lattices, but it isn't one. However, it transforms into a Bravais lattice when the basis is set to two atoms (each lattice point represents one atom). The primitive physical lattice vectors of graphene are defined as follows:
    \begin{gather}
        \Vec{l}_{1} = \frac{\ell}{2} \left( 3\hat{x} + \sqrt{3}\hat{y} \right) \nonumber \\
        \Vec{l}_{2} = \frac{\ell}{2} \left( 3\hat{x} - \sqrt{3}\hat{y} \right) \nonumber \\
        \Vec{\ell} = \ell\hat{x}
        \label{G01}
    \end{gather}
where the last vector is the representation of the basis. These are the reciprocal lattice vectors:
    \begin{gather}
        \Vec{g}_{1} = \frac{2\pi}{3\ell} \left( \hat{k}_{x} + \sqrt{3}\hat{k}_{y} \right) \nonumber \\
        \Vec{g}_{2} = \frac{2\pi}{3\ell} \left( \hat{k}_{x} - \sqrt{3}\hat{k}_{y} \right)
        \label{G02}
    \end{gather}
Also, the reciprocal lattice vectors' sum and difference are given by:
    \begin{gather}
        \Vec{f}_{1} = \Vec{g}_{1} + \Vec{g}_{2} = \frac{4\pi}{3\ell}\hat{k}_{x} \nonumber \\
        \Vec{f}_{2} = \Vec{g}_{1} - \Vec{g}_{2} = \frac{4\pi}{\sqrt{3}\ell}\hat{k}_{y}
        \label{G03}
    \end{gather}
The aforementioned vectors are necessary in the application of the formalism in a graphene lattice.

\subsection{The Hilbert space (physical properties)}
\label{SectG1}

The physical lattice is defined by the following:
    \begin{equation}
        \mathscr{O} \equiv
        \begin{rcases}
            \begin{dcases}
                2c_{1}^{1}\Vec{l}_{1} + 2c_{2}^{1}\Vec{l}_{2}, & c_{1,2}^{1} \in{\mathbb{Z}} \\
                2\Vec{\ell} + 2c_{1}^{2}\Vec{l}_{1} + 2c_{2}^{2}\Vec{l}_{2}, & c_{1,2}^{2} \in{\mathbb{Z}}
            \end{dcases}
        \end{rcases}
        \label{GA01}
    \end{equation}
The physical lattice's first Brillouin zone is given by:
    \begin{equation}
        \mathscr{M} = \left\{ \frac{1}{2}m_{1} \Vec{f}_{1} + \frac{1}{2}m_{2} \Vec{f}_{2},\;
            \begin{dcases}
                m_{1} \in \left( -1/2,1/2 \right] \\
                m_{2} \in \left( -1/4,1/4 \right]
            \end{dcases} \right\}
        \label{GA02}
    \end{equation}

The auxiliary lattice, denoted by the symbol $\mathscr{O'}$ and comprised of three distinct translations of the physical lattice by half of its spacing, is further defined as follows:
    \begin{equation}
        \mathscr{O'} \equiv
        \begin{rcases}
            \begin{dcases}
                \left( 2c_{1}^{1} + 1 \right)\Vec{l}_{1} + 2c_{2}^{1}\Vec{l}_{2} \\
                2\Vec{\ell} + \left( 2c_{1}^{1} + 1 \right)\Vec{l}_{1} + 2c_{2}^{1}\Vec{l}_{2} \\
                2c_{1}^{2}\Vec{l}_{1} + \left( 2c_{2}^{2} + 1 \right)\Vec{l}_{2} \\
                2\Vec{\ell} + 2c_{1}^{2}\Vec{l}_{1} + \left( 2c_{2}^{2} + 1 \right)\Vec{l}_{2} \\
                \left( 2c_{1}^{3} + 1 \right)\Vec{l}_{1} + \left( 2c_{2}^{3} + 1 \right)\Vec{l}_{2} \\
                2\Vec{\ell} + \left( 2c_{1}^{3} + 1 \right) \Vec{l}_{1} + \left( 2c_{2}^{3} + 1 \right)\Vec{l}_{2} \\
            \end{dcases}
        \end{rcases}
        \label{GA03}
    \end{equation}
where $c_{1,2}^{1}, c_{1,2}^{2}, c_{1,2}^{3} \in{\mathbb{Z}}$. This leads to the definition of the extended lattice, which is the union of the physical and auxiliary lattices:
    \begin{equation}
        \mathfrak{D} \equiv
        \begin{rcases}
            \begin{dcases}
                c_{1}^{1}\Vec{l}_{1} + c_{2}^{1}\Vec{l}_{2}, & c_{1,2}^{1} \in{\mathbb{Z}} \\
                -\Vec{\ell} + c_{1}^{2}\Vec{l}_{1} + c_{2}^{2}\Vec{l}_{2}, & c_{1,2}^{2} \in{\mathbb{Z}}
            \end{dcases}
        \end{rcases} = \mathscr{O}\cup\mathscr{O'}
        \label{GA04}
    \end{equation}
The extended lattice's first Brillouin zone is given by:
    \begin{equation}
        \mathfrak{M} = \left\{ m_{1} \Vec{f}_{1} + m_{2} \Vec{f}_{2},\;
            \begin{dcases}
                m_{1} \in \left( -1/2,1/2 \right] \\
                m_{2} \in \left( -1/4,1/4 \right]
            \end{dcases} \right\}
        \label{GA05}
    \end{equation}

The following definitions describe the characteristics of the physical states (physical means defined on the physical lattice $\mathscr{O}$):
    \begin{gather}
        \hat{1}_{\mathscr{O}} = \sum_{x \in \mathscr{O}} |x\rangle \langle x| = \int_{\mathscr{M}} d^{2}p |p\rangle \langle p| \qquad \langle x|p \rangle = \frac{1}{\sqrt{|\mathscr{M}|}}e^{ixp} \nonumber \\
        \langle p|q \rangle = \delta^{\left[ \left( \frac{1}{2} \Vec{g}_{1},\frac{1}{2}\Vec{g}_{2} \right) \right]}(p - q) \qquad \langle x|y \rangle = \delta_{x,y} \nonumber \\
        |p \rangle = \frac{1}{\sqrt{|\mathscr{M}|}}\sum_{x\in\mathscr{O}} e^{ixp}|x\rangle
        \label{GA06}
    \end{gather}
where the latter is the physical Fourier decomposition.

\newpage

\begin{figure}[H]
    \centering
        \begin{minipage}{0.45\textwidth}
            \centering
            \includegraphics[scale = 0.5]{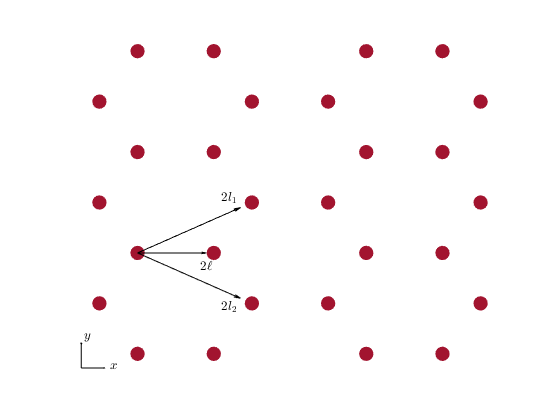}
            \caption{An illustration of the physical lattice $\mathscr{O}$.}
            \label{fig:physical}
        \end{minipage}\hfill
        \begin{minipage}{0.45\textwidth}
            \centering
            \includegraphics[scale = 0.5]{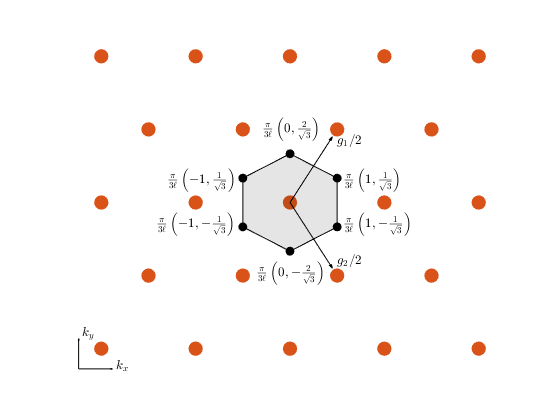}
            \caption{The first Brillouin zone and the reciprocal lattice of $\mathscr{O}$.}
            \label{fig:physical BZ}
        \end{minipage}\hfill
        \begin{minipage}{0.45\textwidth}
            \centering
            \includegraphics[scale = 0.5]{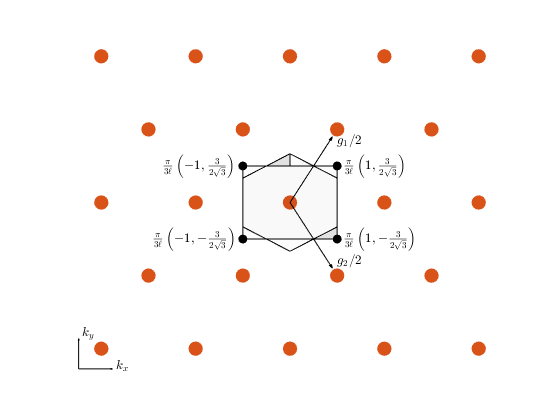}
            \caption{The translation of the first Brillouin zone; it is denoted by the symbol $\mathscr{M}$ and can be seen in (\ref{GA02}); the contributions from shaded areas are equal.}
            \label{fig:translation physical BZ}
        \end{minipage}
\end{figure}

\begin{figure}[H]
    \centering
        \begin{minipage}{0.45\textwidth}
            \centering
            \includegraphics[scale = 0.5]{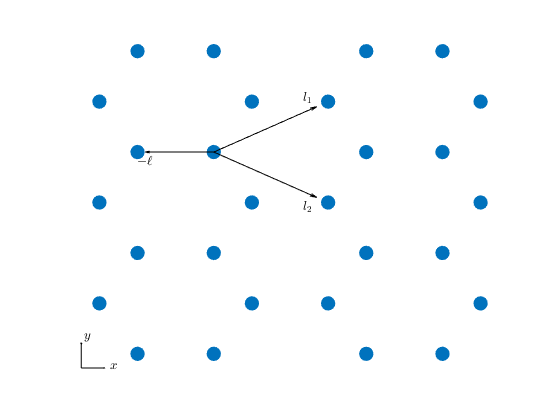}
            \caption{An illustration of the extended lattice $\mathfrak{D}$.}
            \label{fig:extended}
        \end{minipage}\hfill
        \begin{minipage}{0.45\textwidth}
            \includegraphics[scale = 0.5]{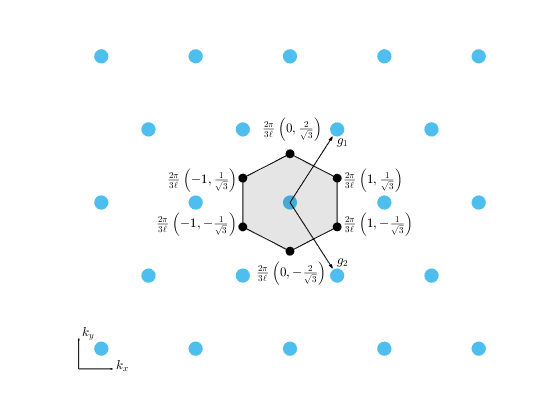}
            \caption{The first Brillouin zone and the reciprocal lattice of $\mathfrak{D}$.}
            \label{fig:extended BZ}
        \end{minipage}\hfill
        \begin{minipage}{0.45\textwidth}
            \includegraphics[scale = 0.5]{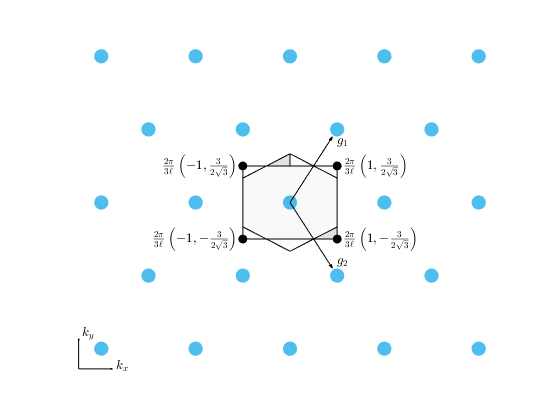}
            \caption{The translation of the first Brillouin zone; it is denoted by the symbol $\mathfrak{M}$ and can be seen in (\ref{GA05}); the contributions from shaded areas are equal.}
            \label{fig:translation extended BZ}
        \end{minipage}
\end{figure}

\newpage

\subsection{$\mathscr{B}$-symbol}
\label{SectG2}

The following definition applies to the two-dimensional $\mathscr{B}$-symbol of an operator $\hat{B}$:
    \begin{align}
        B_{\mathscr{B}}(x,p) &\equiv \int_{\mathscr{M}} d^{2}q e^{2ixq} \langle p + q|\hat{B}|p - q\rangle \nonumber \\
        &= \int_{\mathscr{M}} d^{2}q e^{2ixq}B(p + q,p - q) \nonumber \\
        &= \frac{1}{4} \int_{\mathfrak{M}} d^{2}q e^{ixq}B \left( p + \frac{q}{2},p - \frac{q}{2} \right)
        \label{GB01}
    \end{align}
In contrast to Sect. \ref{Sect2}-(\ref{B01}), a $1/4$ factor is produced when a second dimension is added to the definition of an operator's $\mathscr{B}$-symbol.
Moreover, the following is the equation for the inverse transformation, or $Q(p,q)$, of the function $Q_\mathscr{B}(x,p)$:
    \begin{equation}
        Q(p,q) = \frac{1}{|\mathscr{M}|}\sum_{x \in \mathfrak{D}} e^{-i(p - q)x}Q_{\mathscr{B}} \left( x,\frac{p + q}{2} \right)
        \label{GB02}
    \end{equation}
The following symbols have to be defined in order to show that the inverse transformation holds true in this two-dimensional case:
    \begin{gather}
        \delta^{[a]} (q) = \sum_{n \in \mathbb{Z}} \delta \left[ q - an \right] \nonumber \\
        \Tilde{\delta}^{[a]} (q) = \sum_{n \in \mathbb{Z}} \delta \left[ q - a \Tilde{n} \right]
        \label{GB03}
    \end{gather}
where $\Tilde{n} = n - 1/2$. These Dirac functions $\delta$ and $\Tilde{\delta}$ denote, respectively, even and odd shifts in steps of $\pi$. Therefore, the inverse transformation can be demonstrated as follows:
    \begin{align*}
        &\frac{1}{|\mathscr{M}|} \sum_{x \in \mathfrak{D}} e^{2ix \left(q - \frac{k}{2} \right)} = \frac{1}{|\mathscr{M}|} \sum_{c_{1},c_{2} \in \mathbb{Z}} e^{i\frac{3\ell}{2} (c_{1} + c_{2})(2q_{x} - k_{x})} \nonumber \\
        &\times e^{i\frac{\sqrt{3}\ell}{2} (c_{1} - c_{2})(2q_{y} - k_{y})} \nonumber \\
        &= \frac{1}{|\mathscr{M}|} \sum_{d_{1},d_{2} \in \mathbb{Z}} e^{i\frac{3\ell}{2}(2q_{x} - k_{x})d_{1}} e^{i\frac{\sqrt{3}\ell}{2}(2q_{y} - k_{y})d_{2}} \nonumber \\
        &\times \frac{1}{2} \left( 1 + e^{i\pi(d_{1} + d_{2})} \right) \nonumber \\
        &= \frac{1}{2|\mathscr{M}|} \sum_{d_{1},d_{2} \in \mathbb{Z}} e^{i\frac{3\ell}{2}(2q_{x} - k_{x})d_{1}} e^{i\frac{\sqrt{3}\ell}{2}(2q_{y} - k_{y})d_{2}} \nonumber \\
        &+ \frac{1}{2|\mathscr{M}|} \sum_{d_{1},d_{2} \in \mathbb{Z}} e^{ i\left[ \frac{3\ell}{2}(2q_{x} - k_{x}) + \pi \right]d_{1}} e^{i \left[ \frac{\sqrt{3}\ell}{2}(2q_{y} - k_{y}) + \pi \right]d_{2}} \nonumber \\
        &= \sum_{n \in \mathbb{Z}} \delta \left[ \left( q_{x} - \frac{k_{x}}{2} \right) -\frac{2\pi}{3\ell}n \right] \delta \left[ \left( q_{y} - \frac{k_{y}}{2} \right) - \frac{2\pi}{\sqrt{3}\ell}n \right] \nonumber \\
        &+ \sum_{n \in \mathbb{Z}} \delta \left[\left(q_{x} - \frac{k_{x}}{2} \right) - \frac{2\pi}{3\ell} \Tilde{n} \right] \delta \left[ \left( q_{y} - \frac{k_{y}}{2} \right) -\frac{2\pi}{\sqrt{3}\ell} \Tilde{n} \right] \nonumber \\
        &= \delta^{\left[ \frac{2\pi}{3\ell} \right]} \left( q_{x} - \frac{k_{x}}{2} \right) \delta^{\left[ \frac{2\pi}{\sqrt{3}\ell} \right]}\left( q_{y} - \frac{k_{y}}{2} \right) \nonumber \\
        &+ \Tilde{\delta}^{\left[ \frac{2\pi}{3\ell} \right]} \left( q_{x} - \frac{k_{x}}{2} \right)\Tilde{\delta}^{\left[ \frac{2\pi}{\sqrt{3}\ell} \right]}\left( q_{y} - \frac{k_{y}}{2} \right),
    \end{align*}
such that
    \begin{align*}
        &\frac{1}{|\mathscr{M}|} \sum_{x \in \mathfrak{D}} e^{-ikx}Q_{\mathscr{B}}(x,p) = \frac{1}{|\mathscr{M}|} \sum_{x \in \mathfrak{D}} \int_{\mathscr{M}} d^{2}q e^{2ix \left( q - \frac{k}{2} \right)} \nonumber \\
        &\times Q(p + q,p - q) \nonumber \\
        &= \int_{\mathscr{M}} d^{2}q \delta^{\left[ \frac{1}{2}\Vec{f}_{1},\frac{1}{2}\Vec{f}_{2} \right]} \left( q - \frac{k}{2} \right) Q(p + q,p - q) \nonumber \\
        &+ \int_{\mathscr{M}} d^{2}q \Tilde{\delta}^{\left[ \frac{1}{2}\Vec{f}_{1},\frac{1}{2}\Vec{f}_{2} \right]} \left( q - \frac{k}{2} \right) Q(p + q,p - q) \nonumber \\
        &= \int_{\mathscr{M}} d^{2}q \delta^{\left[ \left( \frac{1}{2}\Vec{g}_{1},\frac{1}{2}\Vec{g}_{2} \right) \right]} \left( q - \frac{k}{2} \right) Q(p + q,p - q) \nonumber \\
        &= Q \left(p + \frac{k}{2},p - \frac{k}{2} \right) \tag*{$\blacksquare$}
    \end{align*}
The Buot symbol of an operator can also be expressed as
 \begin{eqnarray}
	{B}_{\mathscr{B}}(x,p)
	&&	\equiv \int_{ {\cal M}} \D q e^{2\ii x q}\langle {p+q}| \hat{B} |{p-q}\rangle\nonumber\\
	&&	
	= \sum_{z,y\in \mathscr{O}} \int_{\mathscr{M}} \D q e^{2\ii x q} \langle{ {p+q}|z}\rangle\langle{ z| \hat{B}| y}\rangle \langle{ y |{p-q}}\rangle\nonumber \\
	&&
	= \frac{1}{|\mathscr{M}|}\sum_{z,y\in \mathscr{O}} \int_{ \mathscr{M}} \D q e^{2\ii x q -\ii(p+q)z+\ii(p-q)y} \langle{ z| \hat{B}| y}\rangle\nonumber \\
	&&
	= \sum_{z,y\in \mathscr{O}}  e^{-\ii p(z-y)} {\bf d}(2x - z-y) \langle{ z| \hat{B}| y}\rangle,
	\label{GWxAH1}
 \end{eqnarray}
where
$$
{\bf d}(w)
= \frac{1}{|{\cal M}|} \int_{\cM} \D{q} e^{\ii wq}.
$$
Here the integral is over the Brillouin zone. Function $B_{\mathscr{B}}(x,p)$ is defined by \Ref{GWxAH1} for any real-valued $x$, not only for the values of $x\in \mathscr{O}$. Function ${\bf d}(w)$ is reduced to $\delta_{w,0}$ for $w\in \mathscr{O}$.
\subsection{Moyal product}
\label{SectG3}

The two-dimensional case is consistent with the definition of the star-product in Sect. \ref{SectIntro}-(\ref{01}) and the two-dimensional definition of the $\mathscr{B}$-symbol in (\ref{GB01}). Consider the following as proof:
    \begin{align*}
        &A_{\mathscr{B}}(x,p) \star B_{\mathscr{B}}(x,p) = \frac{1}{4} \int_{\mathfrak{M}} d^{2}q e^{ixq} \left< p + \frac{q}{2} \left| \hat{A} \right| p - \frac{q}{2} \right> \nonumber \\
        &\times e^{\frac{i}{2} \left( \overleftarrow{\partial_{x}}\overrightarrow{\partial_{p}} - \overleftarrow{\partial_{p}}\overrightarrow{\partial_{x}} \right)} \times \frac{1}{4} \int_{\mathfrak{M}} d^{2}k e^{ixk} \left< p + \frac{k}{2} \left| \hat{B} \right| p - \frac{k}{2} \right> \nonumber \\
        &= \frac{1}{4^{2}} \int_{\mathfrak{M}} d^{2}q d^{2}k e^{ix(q + k)} \left< p + \frac{q}{2} + \frac{k}{2} \left| \hat{A} \right| p - \frac{q}{2} + \frac{k}{2} \right> \nonumber \\
        &\times \left< p + \frac{k}{2} - \frac{q}{2} \left| \hat{B} \right| p - \frac{k}{2} - \frac{q}{2} \right> \nonumber \\
        &= \frac{2^{2}}{2^{2}\times 4^{2}} \int_{\mathfrak{M}} d^{2}\mathcal{P} d^{2}\mathcal{K} e^{ix \mathcal{P}} \left< p + \frac{\mathcal{P}}{2} \left| \hat{A} \right| p - \frac{\mathcal{K}}{2} \right> \nonumber \\
        &\times \left< p - \frac{\mathcal{K}}{2} \left| \hat{B} \right| p - \frac{\mathcal{P}}{2} \right> \nonumber \\
        &= \frac{1}{4} \int_{\mathfrak{M}} d^{2}\mathcal{P} e^{ix \mathcal{P}} \int_{\mathscr{M}} d^{2}\mathcal{K} \left< p + \frac{\mathcal{P}}{2} \left| \hat{A} \right| p - \mathcal{K} \right> \nonumber \\
        &\times \left< p - \mathcal{K} \left| \hat{B} \right| p - \frac{\mathcal{P}}{2} \right> \nonumber \\
        &= \frac{1}{4} \int_{\mathfrak{M}} d^{2}\mathcal{P} e^{ix\mathcal{P}} \left< p + \frac{\mathcal{P}}{2} \left| \hat{A}\hat{B} \right| p - \frac{\mathcal{P}}{2} \right> \tag*{$\blacksquare$}
     \end{align*}
The transition to the third line is possible via the Taylor expansion given below:
    \begin{equation}
        e^{a\partial_{x}}f(x) = \sum_{n = 0}^{\infty} \frac{a^{n}}{n!}\partial_{x}^{n}f(x) = f(x + a)
        \label{GC01}
    \end{equation}
The origin of the $1/2^{2}$ factor in the fifth line is the Jacobian generated by changing the variables $\mathcal{P} = q + k$ and $\mathcal{K} = q - k$, where $q,k \in \mathfrak{M}$. When the variables are changed, the integration region also changes from a four-dimensional square to a four-dimensional rhomboid. The $x$ and $y$ projections of the four-dimensional square and rhomboid regions are shown in FIGS. \ref{fig:starx} and \ref{fig:stary}, respectively:
\begin{figure}[H]
    \centering
        \begin{minipage}{0.45\textwidth}
            \centering
            \includegraphics[scale = 0.5]{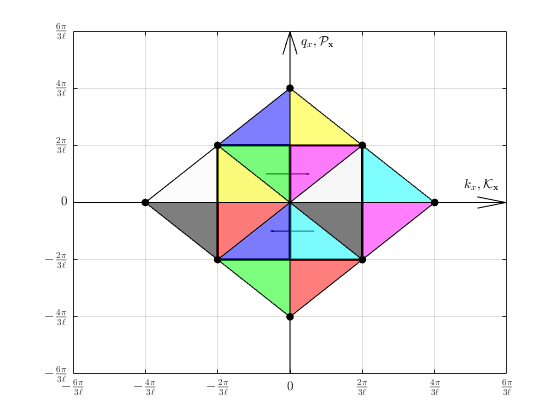}
            \caption{The $x$ projections of the four-dimensional square and rhomboid regions; the contributions from colored areas are equal.}
            \label{fig:starx}
        \end{minipage}\hfill
        \begin{minipage}{0.45\textwidth}
            \centering
            \includegraphics[scale = 0.5]{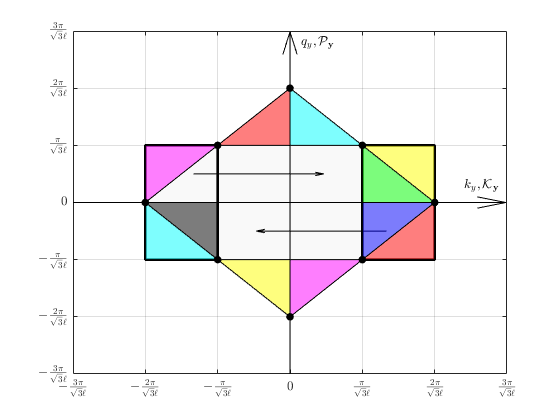}
            \caption{The $y$ projections of the four-dimensional square and rhomboid regions; the contributions from colored areas are equal.}
            \label{fig:stary}
        \end{minipage}
\end{figure}
Due to the integral's periodic nature, with periods of $2\pi/3\ell$ in $x$ and $2\pi/\sqrt{3}\ell$ in $y$, the four-dimensional rhomboid region can be converted back into the four-dimensional square region. The white, grey, green, and blue triangles are shifted twice by $4\pi/3\ell$ in $x$, whereas the red, yellow, cyan, and magenta triangles are shifted once by $\pm 2\pi/3\ell$ in $x$ and once more by $\pm 2\pi/\sqrt{3}\ell$ in $y$. The black-outlined rectangles are eventually shifted once more in $x$ and $y$ (the shifts are denoted by the arrows). This shifts in $x$ and $y$ are linearly correlated, which means there are eight four-dimensional triangles that can be shifted into a single four-dimensional square, making the four-dimensional rhomboid region $2^{2}$ times the size of the four-dimensional square region. Thus, the factor $1/2^{2}$ appears in the fifth line when the integration region is transformed back from a four-dimensional rhomboid to a four-dimensional square.

\newpage

\subsection{Trace and its properties}
\label{SectG4}

The definitions of the traces of the physical and extended lattices are as follows:
    \begin{align}
        &{\rm Tr}_{\mathscr{O}} A_{\mathscr{B}} \equiv \sum_{x \in \mathscr{O}} \int_{\mathscr{M}} \frac{d^{2}p}{|\mathscr{M}|} A_{\mathscr{B}}(x,p) \nonumber \\
        &{\rm Tr}_{\mathfrak{D}} A_{\mathscr{B}} \equiv \sum_{x \in \mathfrak{D}} \int_{\mathscr{M}} \frac{d^{2}p}{|\mathscr{M}|} A_{\mathscr{B}}(x,p)
        \label{GD01}
    \end{align}
Both adhere to the first trace identity stated in Sect. \ref{SectIntro}-(\ref{02}). As proof, consider the following:
    \begin{align*}
        &{\rm Tr}_{\mathfrak{D}}A_{\mathscr{B}} \equiv \frac{1}{|\mathscr{M}|}\sum_{x\in \mathfrak{D}} \int_{\mathscr{M}} d^{2}p A_{\mathscr{B}}(x,p) \nonumber \\
        &= \frac{1}{|\mathscr{M}|}\sum_{x\in \mathfrak{D}} \int_{\mathscr{M}} d^{2}p d^{2}q e^{2ixq} \langle p + q|\hat{A}|p - q \rangle \nonumber \\
        &= \frac{1}{|\mathscr{M}|}\sum_{x\in \mathfrak{D}} \sum_{x,y\in \mathscr{O}} \int_{\mathscr{M}} d^{2}p d^{2}q e^{2ixq} \langle p + q|z \rangle \langle z|\hat{A}|y \rangle \langle y|p - q \rangle \nonumber \\
        &= \frac{1}{|\mathscr{M}|^{2}}\sum_{x\in \mathfrak{D}} \sum_{z,y\in \mathscr{O}} \int_{\mathscr{M}} d^{2}p d^{2}q e^{i(2x - z - y)q} e^{i(y - z)p} \langle z|\hat{A}|y \rangle \nonumber \\
        &= \sum_{x\in \mathfrak{D}} \sum_{y,z\in \mathscr{O}} \delta_{2x,y + z}\delta_{y,z} \langle z|\hat{A}|y \rangle \nonumber \\
        &= \sum_{x\in \mathscr{O}} \langle x|\hat{A}|x \rangle = {\rm tr}\hat{A} \tag*{$\blacksquare$}
    \end{align*}
The same can be demonstrated for the physical lattice, but only the extended lattice accommodates the second trace identity specified in Sect. \ref{SectIntro}-(\ref{03}). The proof:
    \begin{align*}
        &{\rm Tr}_{\mathfrak{D}}(A_{\mathscr{B}} \star B_{\mathscr{B}}) \equiv \sum_{x \in \mathfrak{D}} \int_{\mathscr{M}} \frac{d^{2}p}{|\mathscr{M}|} A_{\mathscr{B}}(x,p) \star B_{\mathscr{B}}(x,p) \nonumber \\
        &= \frac{1}{|\mathscr{M}|} \sum_{x \in \mathfrak{D}} \int_{\mathscr{M}} d^{2}p d^{2}q d^{2}k e^{2ixq} \langle p + q|\hat{A}|p - q \rangle \nonumber \\
        &\times e^{\frac{i}{2} \left( \overleftarrow{\partial_{x}}\overrightarrow{\partial_{p}} - \overleftarrow{\partial_{p}}\overrightarrow{\partial_{x}} \right)} e^{2ixk} \langle p + k|\hat{B}|p - k \rangle \nonumber \\
        &= \frac{1}{|\mathscr{M}|^{3}} \sum_{x \in \mathfrak{D}} \sum_{y,z,\Bar{y},\Bar{z}\in \mathscr{O}} \int_{\mathscr{M}} d^{2}p d^{2}q d^{2}k e^{i(2x - y - z)q + i(z- y)p} \nonumber \\
        &\times \langle y|\hat{A}|z \rangle e^{\frac{i}{2} \left( \overleftarrow{\partial_{x}}\overrightarrow{\partial_{p}} - \overleftarrow{\partial_{p}}\overrightarrow{\partial_{x}} \right)} e^{i(2x - \Bar{y} - \Bar{z})k + i(\Bar{z} - \Bar{y})p} \langle \Bar{y}|\hat{B}|\Bar{z} \rangle \nonumber \\
        &= \frac{1}{|\mathscr{M}|^{3}} \sum_{x \in \mathfrak{D}} \sum_{y,z,\Bar{y},\Bar{z}\in \mathscr{O}} \int_{\mathscr{M}} d^{2}p d^{2}q d^{2}k e^{i(2x - y - z)q + i(z- y)p} \nonumber \\
        &\times \langle y|\hat{A}|z \rangle e^{i(\Bar{y} - \Bar{z})q + i(z - y)k} e^{i(2x - \Bar{y} - \Bar{z})k + i(\Bar{z} - \Bar{y})p} \langle \Bar{y}|\hat{B}|\Bar{z} \rangle \nonumber \\
        &= \sum_{x \in \mathfrak{D}} \sum_{y,z,\Bar{y},\Bar{z} \in \mathscr{O}} \delta_{2x + \Bar{y} - \Bar{z},y + z} \langle y|\hat{A}|z \rangle \delta_{y - z,\Bar{z} - \Bar{y}} \nonumber \\
        &\times \delta_{2x - y + z,\Bar{y} + \Bar{z}} \langle \Bar{y}|\hat{B}|\Bar{z} \rangle \nonumber \\
        &= \sum_{x\in \mathscr{O}} \langle x|\hat{A}\hat{B}|x \rangle = {\rm tr}\hat{A}\hat{B} \tag*{$\blacksquare$}
    \end{align*}

\subsection{$\mathscr{B}$-symbol of the identity operator}
\label{SectG5}

The identity operator's two-dimensional $\mathscr{B}$-symbol is represented by:
    \begin{equation}
        (\hat{1})_{\mathscr{B}}(x,p) = \frac{1}{4} \left[ 1 + \cos{\pi c_{1}^{1}} + \cos{\pi c_{2}^{1}} + \cos{\pi \left( c_{1}^{1} + c_{2}^{1} \right)} \right]
        \label{GE01}
    \end{equation}
Here $x \in \mathfrak{D}, p \in \mathscr{M}$, and
 \begin{equation}
	\mathfrak{D} \equiv
	\begin{rcases}
		\begin{dcases}
			c_{1}^{1}\Vec{l}_{1} + c_{2}^{1}\Vec{l}_{2}, & c_{1,2}^{1} \in{\mathbb{Z}} \\
			-\Vec{\ell} + c_{1}^{2}\Vec{l}_{1} + c_{2}^{2}\Vec{l}_{2}, & c_{1,2}^{2} \in{\mathbb{Z}}
		\end{dcases}
	\end{rcases} = \mathscr{O}\cup\mathscr{O'}
	\label{GA042}
\end{equation}
It is clear that Weyl symbol of identity operator is equal to unity if both $c_{1}^{1}$ and $c_{2}^{1}$ are even integers (i.e. $x \in \mathscr{O}$), but it is not equal to unity if either $c_{1}^{1}$ or $c_{2}^{1}$ is an odd integer. As a result the function $1_W$ instead of being constant is fast oscillating. Take into account the following when validating (\ref{GE01}):
    \begin{align*}
        &(\hat{1})_{\mathscr{B}} = \int_{\mathscr{M}} d^{2}q e^{2ixq}\langle p + q|p - q\rangle \nonumber \\
        &= \int_{\mathscr{M}} d^{2}q e^{2ixq} \delta^{\left[ \left( \frac{1}{2}\Vec{g}_{1},\frac{1}{2}\Vec{g}_{2} \right) \right]}(2q) \nonumber \\
        &= \frac{1}{4} \int_{\mathscr{M}} d^{2}q e^{2ixq} \delta^{\left[ \left( \frac{1}{4}\Vec{g}_{1},\frac{1}{4}\Vec{g}_{2} \right) \right]}(q) \nonumber \\
        &= \frac{1}{4} \int_{-\frac{\pi}{3\ell}}^{\frac{\pi}{3\ell}} dq_{x} e^{2ixq_{x}} \delta \left( q_{x} \right) \int_{-\frac{\pi}{2\sqrt{3}\ell}}^{\frac{\pi}{2\sqrt{3}\ell}} dq_{y} e^{2iyq_{y}} \delta \left( q_{y} \right) \nonumber \\
        &+ \frac{1}{4} \int_{-\frac{\pi}{3\ell}}^{\frac{\pi}{3\ell}} dq_{x} e^{2ixq_{x}} \delta \left( q_{x} \mp \frac{\pi}{6\ell} \right) \int_{-\frac{\pi}{2\sqrt{3}\ell}}^{\frac{\pi}{2\sqrt{3}\ell}} dq_{y} e^{2iyq_{y}} \nonumber \\
        &\times \delta \left( q_{y} \pm \frac{\pi}{2\sqrt{3}\ell} \right) + \frac{1}{4} \int_{-\frac{\pi}{3\ell}}^{\frac{\pi}{3\ell}} dq_{x} e^{2ixq_{x}} \delta \left( q_{x} \pm \frac{\pi}{6\ell} \right) \nonumber \\
        &\times \int_{-\frac{\pi}{2\sqrt{3}\ell}}^{\frac{\pi}{2\sqrt{3}\ell}} dq_{y} e^{2iyq_{y}} \delta \left( q_{y} \pm \frac{\pi}{2\sqrt{3}\ell} \right) + \frac{1}{4} \int_{-\frac{\pi}{3\ell}}^{\frac{\pi}{3\ell}} dq_{x} e^{2ixq_{x}} \nonumber \\
        &\times \delta \left( q_{x} \pm \frac{\pi}{3\ell} \right) \int_{-\frac{\pi}{2\sqrt{3}\ell}}^{\frac{\pi}{2\sqrt{3}\ell}} dq_{y} e^{2iyq_{y}} \delta \left( q_{y} \right) \nonumber \\
        &= \frac{1}{4} + \frac{1}{8}e^{i\frac{\pi}{3\ell}x}e^{-i\frac{\pi}{\sqrt{3}\ell}y} + \frac{1}{8}e^{-i\frac{\pi}{3\ell}x}e^{i\frac{\pi}{\sqrt{3}\ell}y} + \frac{1}{8}e^{-i\frac{\pi}{3\ell}x}e^{-i\frac{\pi}{\sqrt{3}\ell}y} \nonumber \\
        &+ \frac{1}{8}e^{i\frac{\pi}{3\ell}x}e^{i\frac{\pi}{\sqrt{3}\ell}y} + \frac{1}{8}e^{-i\frac{2\pi}{3\ell}x} + \frac{1}{8}e^{i\frac{2\pi}{3\ell}x} \nonumber \\
        &= \frac{1}{4} + \frac{1}{2}\cos{\frac{\pi}{3\ell}x}\cos{\frac{\pi}{\sqrt{3}\ell}y} + \frac{1}{4}\cos{\frac{2\pi}{3\ell}x},
    \end{align*}
where
    \begin{gather*}
        x = \left( c_{1}^{1}\Vec{l}_{1} + c_{2}^{1}\Vec{l}_{2} \right)_{x} = \frac{3\ell}{2} \left( c_{1}^{1} + c_{2}^{1} \right) \nonumber \\
        y = \left( c_{1}^{1}\Vec{l}_{1} + c_{2}^{1}\Vec{l}_{2} \right)_{y} = \frac{\sqrt{3}\ell}{2} \left( c_{1}^{1} - c_{2}^{1} \right) \tag*{$\blacksquare$}
    \end{gather*}
These variables, $x$ and $y$, are the linear projections of the extended lattice vectors (lattice $\mathfrak{D}$) onto the axes.

\subsection{Star product without differentiation}

\label{SectBuoStar}
Let us represent the star product of Buot symbols for $x \in \mathscr{D}$ through the matrix elements of $\hat A$ and $\hat B$:
\begin{eqnarray}
	&&A_{\mathscr{B}}(x,p) \star B_{\mathscr{B}}(x,p) =  \int_{\cal M} dk dq e^{\ii  2x k } \langle p+k|  \hat{A}|p-k\rangle \nonumber\\&& \star e^{2\ii  x q }\langle p+q|  \hat{B}|p-q\rangle \nonumber\\
	& = & \frac{1}{|{\cal M}|^2}\sum_{z_1,z_2,\bar{z}_1,\bar{z}_2\in \cO}\int_{\cal M} dk dq e^{\ii  2x k -iz_1(p+k) + iz_2(p-k)} \langle z_1|  \hat{A}|z_2\rangle \star \nonumber\\&& e^{2\ii  x q - i\bar{z}_1(p+q) + i\bar{z}_2(p-q)}\langle \bar{z}_1|  \hat{B}|\bar{z}_2\rangle
	\nonumber\\
	& = & \frac{1}{|{\cal M}|^2} \sum_{z_1,z_2,\bar{z}_1,\bar{z}_2\in \cO}\int_{\cal M} dk dq e^{\ii  (2x-z_1-z_2) k +\ii p(-z_1 + z_2)} \langle z_1|  \hat{A}|z_2\rangle \star \nonumber\\&& e^{\ii  (2x -\bar{z}_1-\bar{z}_2) q +\ii p(-\bar{z}_1+\bar{z}_2) }\langle \bar{z}_1|  \hat{B}|\bar{z}_2\rangle
	\nonumber\\
	& = & \sum_{z_1,z_2,\bar{z}_1,\bar{z}_2\in \cO} \delta_{2x,z_1+z_2} e^{ ip(-z_1 + z_2)} \langle z_1|  \hat{A}|z_2\rangle \star \nonumber\\&& e^{\ii p(-\bar{z}_1+\bar{z}_2)}\delta_{2x,  \bar{z}_1+\bar{z}_2}\langle \bar{z}_1|  \hat{B}|\bar{z}_2\rangle
	\nonumber\\
	& = &  \sum_{z_1,z_2,\bar{z}_1,\bar{z}_2\in \cO} \delta_{2x,z_1+z_2} e^{ i(p-i\vec{\partial_x}/2)(-z_1 + z_2)} \langle z_1|  \hat{A}|z_2\rangle  \nonumber\\&& e^{\ii (p+i\cev{\partial_x}/2)(-\bar{z}_1+\bar{z}_2)}\delta_{2x,  \bar{z}_1+\bar{z}_2}\langle \bar{z}_1|  \hat{B}|\bar{z}_2\rangle
	\nonumber\\
	& = & \sum_{z_1,z_2,\bar{z}_1,\bar{z}_2\in \cO} \delta_{2x+\bar{z}_1-\bar{z}_2,z_1+z_2} e^{ ip(-z_1 + z_2)} \langle z_1|  \hat{A}|z_2\rangle  \nonumber\\&& e^{\ii p(-\bar{z}_1+\bar{z}_2)}\delta_{2x - z_1 + z_2,  \bar{z}_1+\bar{z}_2}\langle \bar{z}_1|  \hat{B}|\bar{z}_2\rangle
\end{eqnarray}
Next, using Eq. (\ref{GWxAH1}) we obtain (for $x \in \mathscr{D}$):
\begin{eqnarray}
	&&A_{\mathscr{B}}(x,p) \star B_{\mathscr{B}}(x,p)
	 =  \sum_{2z,2\bar{z},u,\bar{u} \in \cO} \delta_{2x-\bar{u},2z} \delta_{2x+{u},2\bar{z}}\nonumber\\&&\int \frac{\D{p}^\prime}{|{\cal M}|}\frac{d\bar{p}^\prime}{|{\cal M}|}e^{ ip^\prime u + i \bar{p}^\prime \bar{u}} A_{\mathscr{B}}(z,p-p^\prime)B_{\mathscr{B}}(\bar{z},p-\bar{p}^\prime)\label{starlattice}\label{BBns}
\end{eqnarray}
One can see, that in order to define the star product of the  symbols $A_{\mathscr{B}}(x,p)$ and $B_{\mathscr{B}}(x,p)$ for $x \in \mathscr{D}$ we do not need to know the values of these functions for all real values of $x$. It is enough to know the values of the Weyl symbols for $x \in \mathscr{D}$.



\section{Precise Wigner-Weyl calculus in graphene}
\label{SectPrecise}

Here, the precise Wigner-Weyl formalism—which makes use of the extended lattice—is constructed. We use this precise Wigner-Weyl calculus on the extended lattice of graphene, utilizing both the physical lattice $\mathscr{O}$ and the extended lattice $\mathfrak{D}$, along with their respective first Brillouin zones $\mathscr{M}$ and $\mathfrak{M}$.

\subsection{The Hilbert space (extended properties)}
\label{SectG6}

This section is the same as Sect. \ref{SectG1}, with the exception that the properties of the physical states that act on the physical lattice $\mathscr{O}$ are reconfigured to act on the extended lattice $\mathfrak{D}$. Consequently, the following definitions are given for the extended states' properties:
    \begin{gather}
        \hat{1}_{\mathfrak{D}} = \sum_{x \in \mathfrak{D}} |x \rangle \langle x| = \int_{\mathfrak{M}} d^{2}p |p \rangle \langle p| \qquad \langle x|p \rangle = \frac{1}{\sqrt{|\mathfrak{M}|}}e^{ixp} \nonumber \\
        \langle p|q \rangle = \delta^{\left[ \left( \Vec{g}_{1},\Vec{g}_{2} \right) \right]}(p - q) \qquad \langle x|y \rangle = \delta_{x,y}
        \label{GF01}
    \end{gather}
together with the Fourier decomposition, which is defined as follows:
    \begin{align}
        |p \rangle &= \frac{1}{\sqrt{|\mathfrak{M}|}}\sum_{x\in\mathfrak{D}} e^{ixp}|x \rangle \nonumber \\
        &= \frac{1}{\sqrt{|\mathfrak{M}|}}\sum_{x\in\mathscr{O}} e^{ixp}|x \rangle + \frac{1}{\sqrt{|\mathfrak{M}|}}\sum_{x'\in\mathscr{O'}} e^{ix'p}|x' \rangle
    \end{align}

Additionally, the following relationships hold true for the extended operators:
    \begin{align}
        \langle x|\hat{Q}|y \rangle &= \langle x + l_{1}|\hat{Q}|y + l_{1} \rangle \nonumber \\
        &= \langle x + l_{2}|\hat{Q}|y + l_{2} \rangle \nonumber \\
        &= \langle x + l_{1} + l_{2}|\hat{Q}|y + l_{1} + l_{2} \rangle
        \label{GF02}
    \end{align}
where $x,y\in \mathscr{O}$, and the inter-lattice matrix elements vanish, such that
    \begin{equation}
        \langle x |\hat{Q}| y\rangle = \langle y |\hat{Q}| x\rangle = 0
        \label{GF03}
    \end{equation}
Here, $x\in\mathscr{O}$, and $y\in\mathscr{O'}$. Therefore, the matrix elements in the momentum space of such operators are given by:
    \begin{align}
        \langle p|\hat{Q}|q \rangle &= \frac{1}{4} Q(p,q) \nonumber \\
        &\times \left( 1 + e^{il_{1}(q - p)} + e^{il_{2}(q - p)} + e^{il_{1,2}(q - p)} \right)
        \label{GF04}
    \end{align}
where $l_{1,2} = l_{1} + l_{2}$. It is required to mention the following definition before verifying (\ref{GF04}):
    \begin{equation}
        Q(p,q) = \frac{1}{|\mathscr{M}|} \sum_{x_{1},x_{2} \in \mathscr{O}} \langle x_{1}|\hat{Q}|x_{2} \rangle e^{i(x_{2}q - x_{1}p)}
        \label{GF05}
    \end{equation}
As a result, it is now possible to demonstrate that (\ref{GF04}) is true as follows:
    \begin{align*}
        &\langle p|\hat{Q}|q \rangle = \frac{1}{|\mathfrak{M}|} \sum_{x_{1},x_{2} \in \mathscr{O}} \langle x_{1}|\hat{Q}|x_{2} \rangle e^{i(x_{2}q - x_{1}p)} \nonumber \\
        &+ \frac{1}{|\mathfrak{M}|} \sum_{x'_{1},x'_{2} \in \mathscr{O'}} \langle x'_{1}|\hat{Q}|x'_{2} \rangle e^{i(x'_{2}q - x'_{1}p)} \nonumber \\
        &= \frac{1}{|\mathfrak{M}|} \sum_{x_{1},x_{2} \in \mathscr{O}} \langle x_{1}|\hat{Q}|x_{2} \rangle e^{i(x_{2}q - x_{1}p)} \nonumber \\
        &+ \frac{1}{|\mathfrak{M}|} \sum_{x_{1},x_{2} \in \mathscr{O}} \langle x_{1} + l_{1}|\hat{Q}|x_{2} + l_{1} \rangle e^{i(x_{2}q - x_{1}p)}e^{il_{1}(q - p)} \nonumber \\
        &+ \frac{1}{|\mathfrak{M}|} \sum_{x_{1},x_{2} \in \mathscr{O}} \langle x_{1} + l_{2}|\hat{Q}|x_{2} + l_{2} \rangle e^{i(x_{2}q - x_{1}p)}e^{il_{2}(q - p)} \nonumber \\
        &+ \frac{1}{|\mathfrak{M}|} \sum_{x_{1},x_{2} \in \mathscr{O}} \langle x_{1} + l_{1} + l_{2}|\hat{Q}|x_{2} + l_{1} + l_{2} \rangle e^{i(x_{2}q - x_{1}p)} \nonumber \\
        &\times e^{i(l_{1} + l_{2})(q - p)} \nonumber \\
        &= \frac{1}{|\mathfrak{M}|} \sum_{x_{1},x_{2} \in \mathscr{O}} \langle x_{1}|\hat{Q}|x_{2} \rangle e^{i(x_{2}q - x_{1}p)} \nonumber \\
        &\times \left( 1 + e^{il_{1}(q - p)} + e^{il_{2}(q - p)} + e^{i(l_{1} + l_{2})(q - p)} \right) \tag*{$\blacksquare$}
    \end{align*}
where, in this two-dimensional case, $|\mathscr{M}| = |\mathfrak{M}|/4$. According to that definition, the matrix element $\langle p|\hat{Q}|q \rangle$ is periodic with periods of $2\pi/3\ell$ in $x$ and $2\pi/\sqrt{3}\ell$ in $y$, whereas the function $Q(p,q)$ is periodic with periods of $\pi/3\ell$ in $x$ and $\pi/\sqrt{3}\ell$ in $y$.

\subsection{$W$-symbol}
\label{SectG7}

The $\mathscr{B}$-symbol of an operator, defined on the extended lattice $\mathfrak{D}$, can be used to determine the $W$-symbol of an operator, defined on the physical lattice $\mathscr{O}$. Take into account the following:
    \begin{align}
        Q_{W}(x,p) &\equiv \int_{\mathfrak{M}} d^{2}q e^{2ixq} \langle p + q|\hat{Q}|p - q\rangle \nonumber \\
        &= \frac{1}{4} \int_{\mathfrak{M}} d^{2}q e^{2ixq} Q(p + q,p - q) \nonumber \\
        &\times \left( 1 + e^{-2il_{1}q} + e^{-2il_{2}q} + e^{-2i(l_{1} + l_{2})q} \right) \nonumber \\
        &= \int_{\mathscr{M}} d^{2}q e^{2ixq} Q(p + q,p - q) \nonumber \\
        &\times \left( 1 + e^{-2il_{1}q} + e^{-2il_{2}q} + e^{-2i(l_{1} + l_{2})q} \right)
        \label{GG01}
    \end{align}
where $Q_{W}(x,p)$ is defined for any $x \in \mathbb{R}$, but due to the discrete values of $x \in \mathfrak{D}$ the integration is reduced from $\mathfrak{M}$ to $\mathscr{M}$. In addition, for brevity's sake, the expression in parentheses will henceforth be abbreviated as $f(q)$. Furthermore, the inverse transformation of a function $Q_{W}(x,p)$ is given by:
    \begin{equation}
        Q(p,q) = \frac{f^{-1} \left( \frac{p - q}{2} \right)}{|\mathscr{M}|} \sum_{x \in \mathfrak{D}} e^{-i(p - q)x}Q_{W} \left( x,\frac{p + q}{2} \right)
        \label{GG02}
    \end{equation}
It is possible to validate (\ref{GG02}) in a manner similar to how it was demonstrated in Sect. \ref{SectG2} by doing the following:
    \begin{align*}
        &\frac{1}{|\mathscr{M}|} \sum_{x \in \mathfrak{D}} e^{2ix \left(q - \frac{k}{2} \right)} = \frac{1}{|\mathscr{M}|} \sum_{c_{1},c_{2} \in \mathbb{Z}} e^{i\frac{3\ell}{2} (c_{1} + c_{2})(2q_{x} - k_{x})} \nonumber \\
        &\times e^{i\frac{\sqrt{3}\ell}{2} (c_{1} - c_{2})(2q_{y} - k_{y})} \nonumber \\
        &= \frac{1}{|\mathscr{M}|} \sum_{d_{1},d_{2} \in \mathbb{Z}} e^{i\frac{3\ell}{2}(2q_{x} - k_{x})d_{1}} e^{i\frac{\sqrt{3}\ell}{2}(2q_{y} - k_{y})d_{2}} \nonumber \\
        &\times \frac{1}{2} \left( 1 + e^{i\pi(d_{1} + d_{2})} \right) \nonumber \\
        &= \frac{1}{2|\mathscr{M}|} \sum_{d_{1},d_{2} \in \mathbb{Z}} e^{i\frac{3\ell}{2}(2q_{x} - k_{x})d_{1}} e^{i\frac{\sqrt{3}\ell}{2}(2q_{y} - k_{y})d_{2}} \nonumber \\
        &+ \frac{1}{2|\mathscr{M}|} \sum_{d_{1},d_{2} \in \mathbb{Z}} e^{ i\left[ \frac{3\ell}{2}(2q_{x} - k_{x}) + \pi \right]d_{1}} e^{i \left[ \frac{\sqrt{3}\ell}{2}(2q_{y} - k_{y}) + \pi \right]d_{2}} \nonumber \\
        &= \sum_{n \in \mathbb{Z}} \delta \left[ \left( q_{x} - \frac{k_{x}}{2} \right) -\frac{2\pi}{3\ell}n \right] \delta \left[ \left( q_{y} - \frac{k_{y}}{2} \right) - \frac{2\pi}{\sqrt{3}\ell}n \right] \nonumber \\
        &+ \sum_{n \in \mathbb{Z}} \delta \left[\left(q_{x} - \frac{k_{x}}{2} \right) - \frac{2\pi}{3\ell} \Tilde{n} \right] \delta \left[ \left( q_{y} - \frac{k_{y}}{2} \right) -\frac{2\pi}{\sqrt{3}\ell} \Tilde{n} \right] \nonumber \\
        &= \delta^{\left[ \frac{2\pi}{3\ell} \right]} \left( q_{x} - \frac{k_{x}}{2} \right) \delta^{\left[ \frac{2\pi}{\sqrt{3}\ell} \right]}\left( q_{y} - \frac{k_{y}}{2} \right) \nonumber \\
        &+ \Tilde{\delta}^{\left[ \frac{2\pi}{3\ell} \right]} \left( q_{x} - \frac{k_{x}}{2} \right)\Tilde{\delta}^{\left[ \frac{2\pi}{\sqrt{3}\ell} \right]}\left( q_{y} - \frac{k_{y}}{2} \right),
    \end{align*}
such that
    \begin{align*}
        &\frac{1}{|\mathscr{M}|} \sum_{x \in \mathfrak{D}} e^{-ikx}Q_{W}(x,p) = \frac{1}{|\mathscr{M}|} \sum_{x \in \mathfrak{D}} \int_{\mathscr{M}} d^{2}q e^{2ix \left( q - \frac{k}{2} \right)} \nonumber \\
        &\times Q(p + q,p - q) f(q) \nonumber \\
        &= \int_{\mathscr{M}} d^{2}q \delta^{\left[ \frac{1}{2}\Vec{f}_{1},\frac{1}{2}\Vec{f}_{2} \right]} \left( q - \frac{k}{2} \right) Q(p + q,p - q) f(q) \nonumber \\
        &+ \int_{\mathscr{M}} d^{2}q \Tilde{\delta}^{\left[ \frac{1}{2}\Vec{f}_{1},\frac{1}{2}\Vec{f}_{2} \right]} \left( q - \frac{k}{2} \right) Q(p + q,p - q) f(q) \nonumber \\
        &= \int_{\mathscr{M}} d^{2}q \delta^{\left[ \left( \frac{1}{2}\Vec{g}_{1},\frac{1}{2}\Vec{g}_{2} \right) \right]} \left( q - \frac{k}{2} \right) Q(p + q,p - q) f(q) \nonumber \\
        &= Q \left(p + \frac{k}{2},p - \frac{k}{2} \right) f\left( \frac{k}{2} \right) \tag*{$\blacksquare$}
    \end{align*}

The Buot symbol on the extended lattice may be expressed as
\begin{eqnarray}
	{Q}_{W}(x,p)
	= \sum_{z,y\in \mathscr{D}}  e^{-\ii p(z-y)} {\bf d}(2x - z-y) \langle{ z| \hat{Q}| y}\rangle,
	\label{GWxAH12}
\end{eqnarray}
where
$$
{\bf d}(w)
= \frac{1}{|\mathfrak{M}|} \int_{\mathfrak{M}} \D{q} e^{\ii wq}.
$$
The doubly extended lattice $\mathscr{S}$ may be defined as an extension of $\mathscr{D}$ in the same way as $\mathscr{D}$ is an extension of $\mathscr{O}$.

We should take into account the constraints of Eq. (\ref{GF02}) and Eq. (\ref{GF03}). This results in
\begin{eqnarray}
	{Q}_{W}(x,p)
	&=& \sum_{z,y\in \mathscr{O}; n_i = 0,1}  e^{-\ii p(z-y)}\nonumber\\&& {\bf d}(2x - z-y - 2 l_i n_i) \langle{ z| \hat{Q}| y}\rangle,
	\label{GWxAH13}
\end{eqnarray}
and
\begin{equation}
	{Q}_{W}(x,p)\Big|_{x
		\in \mathscr{S}\setminus \mathscr{D}}=0\label{BPS0}
\end{equation}

\subsection{Moyal product}
\label{SectG8}

\begin{figure}[H]
    \centering
        \begin{minipage}{0.45\textwidth}
            \centering
            \includegraphics[scale = 0.5]{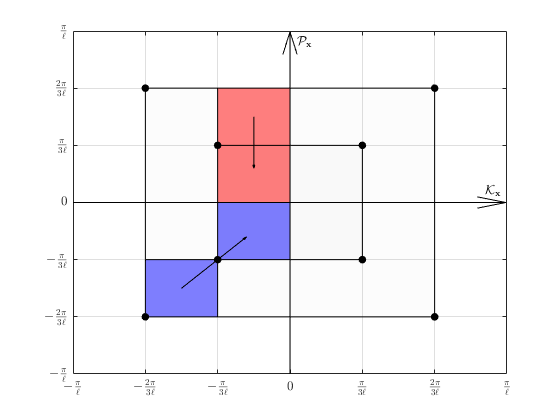}
            \caption{The four-dimensional squares $\mathfrak{M} \otimes \mathfrak{M}$ and $\mathscr{M} \otimes \mathscr{M}$'s $x$ projections. Arrows depict the shifts; red shifts are negative, while blue shifts are positive.}
            \label{fig:Wstarx}
        \end{minipage}\hfill
        \begin{minipage}{0.45\textwidth}
            \centering
            \includegraphics[scale = 0.5]{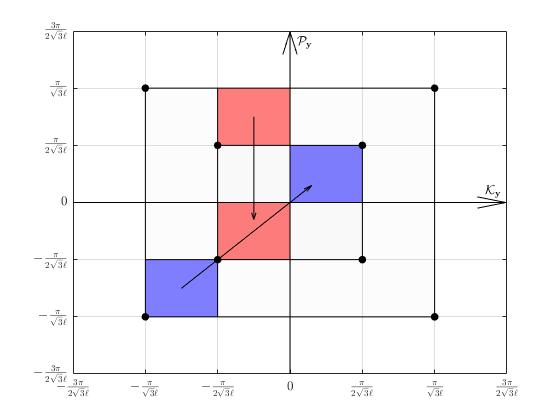}
            \caption{The four-dimensional squares $\mathfrak{M} \otimes \mathfrak{M}$ and $\mathscr{M} \otimes \mathscr{M}$'s $y$ projections. Arrows depict the shifts; red shifts are negative, while blue shifts are positive.}
            \label{fig:Wstary}
        \end{minipage}
\end{figure}

The two-dimensional case is consistent with the definition of the star-product in Sect. \ref{SectIntro}-(\ref{01}) and the definition of the $W$-symbol in (\ref{GG01}). Consider the following as proof:
    \begin{align*}
        &A_{W}(x,p) \star B_{W}(x,p) = \int_{\mathscr{M}} d^{2}q e^{2ixq} f(q) A \left( p + q,p - q \right) \nonumber \\
        &\times e^{\frac{i}{2}(\overleftarrow{\partial_{x}}\overrightarrow{\partial_{p}} - \overleftarrow{\partial_{p}}\overrightarrow{\partial_{x}})} \int_{\mathscr{M}} d^{2}k e^{2ixk} f(k) B \left( p + k,p - k \right) \nonumber \\
        &= \int_{\mathscr{M}} d^{2}q e^{2ixq} f(q) A(p + q + k,p - q + k) \nonumber \\
        &\times \int_{\mathscr{M}} d^{2}k e^{2ixk} f(k) B(p - q + k,p - q - k) \nonumber \\
        &= \frac{1}{2^{2} \times 2^{2}} \int_{\mathfrak{M}} d^{2}\mathcal{P} d^{2}\mathcal{K} e^{2ix\mathcal{P}} f\left( \frac{\mathcal{P} + \mathcal{K}}{2} \right) f\left( \frac{\mathcal{P} - \mathcal{K}}{2} \right) \nonumber \\
        &\times A \left( p + \mathcal{P},p - \mathcal{K} \right) B \left( p - \mathcal{K},p - \mathcal{P} \right) \nonumber \\
        &= \frac{1}{2^{2} \times 2^{2}} \int_{\mathfrak{M}} d^{2}\mathcal{P} d^{2}\mathcal{K} e^{2ix\mathcal{P}} \left[ 1 + g\left( \frac{\mathcal{P} + \mathcal{K}}{2} \right) \right] \nonumber \\
        &\times \left[ 1 + g\left( \frac{\mathcal{P} - \mathcal{K}}{2} \right) \right] A \left( p + \mathcal{P},p - \mathcal{K} \right) B \left( p - \mathcal{K},p - \mathcal{P} \right) \nonumber \\
        &= \frac{1}{2^{2} \times 2^{2}} \int_{\mathfrak{M}} d^{2}\mathcal{P} d^{2}\mathcal{K} e^{2ix\mathcal{P}} \bigg[ 1 + g\left( \frac{\mathcal{P} + \mathcal{K}}{2} \right) + g\left( \frac{\mathcal{P} - \mathcal{K}}{2} \right) \nonumber \\
        &+ g\left( \frac{\mathcal{P} + \mathcal{K}}{2} \right) g\left( \frac{\mathcal{P} - \mathcal{K}}{2} \right) \bigg] A \left( p + \mathcal{P},p - \mathcal{K} \right) \nonumber \\
        &\times B \left( p - \mathcal{K},p - \mathcal{P} \right) \nonumber \\
        &= \frac{4^{2}}{2^{2} \times 2^{2}} \int_{\mathscr{M}} d^{2}\mathcal{P} d^{2}\mathcal{K} e^{2ix\mathcal{P}} f(\mathcal{P}) A \left( p + \mathcal{P},p - \mathcal{K} \right) \nonumber \\
        &\times B \left( p - \mathcal{K},p - \mathcal{P} \right) = (AB)_{W}(x,p) \tag*{$\blacksquare$}
    \end{align*}
The two $1/2^{2}$ factors in the fifth line are produced by the Jacobian and the transition of the integration region from a four-dimensional rhomboid to a four-dimensional square $\mathfrak{M} \otimes \mathfrak{M}$. Due to the periodic nature of the functions $f((\mathcal{P} \pm \mathcal{K})/2)$, which have periods of $2\pi/3\ell$ in $x$ and $2\pi/\sqrt{3}\ell$ in $y$, it is impossible to immediately reduce the integration region from $\mathfrak{M} \otimes \mathfrak{M}$ to $\mathscr{M} \otimes \mathscr{M}$. As a result, the following modification is implemented:
    \begin{equation}
        g(p) \equiv f(p) - 1 = e^{-2il_{1}q} + e^{-2il_{2}q} + e^{-2i(l_{1} + l_{2})q}
    \end{equation}
Therefore, after the expressions in square parentheses are multiplied and the integration region is reduced, the integrals, including the terms $g((\mathcal{P} \pm \mathcal{K})/2)$, cancel out because these functions change sign when shifted by $\pi/3\ell$ in $x$ and/or by $\pi/\sqrt{3}\ell$ in $y$. Those shifts are demonstrated in FIGS. \ref{fig:Wstarx} and \ref{fig:Wstary}. Also canceling out are the integrals, including the terms $1$ and $g((\mathcal{P} + \mathcal{K})/2) \times g((\mathcal{P} - \mathcal{K})/2)$, which don't add up to $f(\mathcal{P})$. Once $f(\mathcal{P})$ is all that is left, the integration region can be reduced from $\mathfrak{M} \otimes \mathfrak{M}$ to $\mathscr{M} \otimes \mathscr{M}$ because it is periodic with periods of $\pi/3\ell$ in $x$ and $\pi/\sqrt{3}\ell$ in $y$. The factor $4^{2}$ in the final line is produced by this reduction.

\newpage

\subsection{Trace and its properties}
\label{SectG9}

The definitions of the traces of the physical and extended lattices are as follows:
    \begin{gather}
        {\rm Tr}_{\mathscr{O}} Q_{W} \equiv \sum_{x \in \mathscr{O}} \int_{\mathscr{M}} \frac{d^{2}p}{|\mathscr{M}|} Q_{W}(x,p) \nonumber \\
        {\rm Tr}_{\mathfrak{D}} Q_{W} \equiv \sum_{x \in \mathfrak{D}} \int_{\mathscr{M}} \frac{d^{2}p}{|\mathfrak{M}|} Q_{W}(x,p)
        \label{GI01}
    \end{gather}
Both adhere to the first trace identity stated in Sect. \ref{SectIntro}-(\ref{02}). As proof, consider the following:
    \begin{align*}
        &{\rm Tr}_{\mathscr{O}} Q_{W} \equiv \sum_{x\in \mathscr{O}} \int_{\mathscr{M}} \frac{d^{2}p}{|\mathscr{M}|} Q_{W}(x,p) \nonumber \\
        &= \frac{1}{|\mathscr{M}|} \sum_{x\in \mathscr{O}} \int_{\mathscr{M}} d^{2}p d^{2}q e^{2ixq} Q(p + q,p - q) f(q) \nonumber \\
        &=\int_{\mathscr{M}} d^{2}p d^{2}q \delta^{\left[ \left( \frac{1}{2}\Vec{g}_{1},\frac{1}{2}\Vec{g}_{2} \right) \right]}(2q) Q(p + q,p - q) f(q) \nonumber \\
        &= \frac{1}{4} \int_{\mathscr{M}} d^{2}p d^{2}q \delta^{\left[ \left( \frac{1}{4}\Vec{g}_{1},\frac{1}{4}\Vec{g}_{2} \right) \right]}(q) Q(p + q,p - q) \nonumber \\
        &\times \left( 1 + e^{-2il_{1}q} + e^{-2il_{2}q} + e^{-2i(l_{1} + l_{2})q} \right) \nonumber \\
        &= \frac{1}{4} \int_{\mathscr{M}} d^{2}p \int_{-\frac{\pi}{3\ell}}^{\frac{\pi}{3\ell}}dq_{x} \int_{-\frac{\pi}{2\sqrt{3}\ell}}^{\frac{\pi}{2\sqrt{3}\ell}}dq_{y} Q(p + q,p - q) \nonumber \\
        &\times \bigg( \delta(q_{x})\delta(q_{y}) + \delta \left( q_{x} \mp \frac{\pi}{6\ell} \right) \delta \left( q_{y} \pm \frac{\pi}{2\sqrt{3}\ell} \right) \nonumber \\
        &+ \delta \left( q_{x} \pm \frac{\pi}{6\ell} \right) \delta \left( q_{y} \pm \frac{\pi}{2\sqrt{3}\ell} \right) + \delta \left( q_{x} \pm \frac{\pi}{3\ell} \right) \delta(q_{y}) \bigg) \nonumber \\
        &\times \bigg( 1 + 2e^{-i3\ell q_{x}}\cos{\sqrt{3}\ell q_{y}} + e^{-i6\ell q_{x}} \bigg) \nonumber \\
        &= \int_{\mathscr{M}} d^{2}p d^{2}q \delta(q) Q(p + q,p - q) \nonumber \\
        &= \int_{\mathscr{M}} d^{2}p \; Q(p,p) = {\rm tr}\hat{Q} \tag*{$\blacksquare$}
    \end{align*}
The same can be demonstrated for the extended lattice. Therefore, both definitions of the traces in (\ref{GI01}) obey the first trace identity as follows:
    \begin{equation}
        {\rm Tr}_{\mathscr{O}} Q_{W} = {\rm Tr}_{\mathfrak{D}} Q_{W} = {\rm tr}\hat{Q}
    \end{equation}

\newpage

The second trace identity stated in Sect. \ref{SectIntro}-(\ref{03}) is solely accommodated by the extended lattice, much like in Sect. \ref{SectG4}. The proof is:
    \begin{align*}
        &{\rm Tr}_{\mathfrak{D}} \left( A_{W}B_{W} \right) \equiv \sum_{x \in \mathfrak{D}} \int_{\mathscr{M}} \frac{d^{2}p}{|\mathfrak{M}|} A_{W}(x,p)B_{W}(x,p) \nonumber \\
        &= \sum_{x \in \mathfrak{D}} \int_{\mathscr{M}} \frac{d^{2}p}{|\mathfrak{M}|} \int_{\mathscr{M}} d^{2}q e^{2ixq} A(p - q,p + q) f(q) \nonumber \\
        &\times \int_{\mathscr{M}} d^{2}k e^{2ixk} B(p - k,p + k) f(k) \nonumber \\
        &= \frac{1}{4} \int_{\mathscr{M}} d^{2}p d^{2}q A(p - q,p + q) B(p + q,p - q) f(q)f(-q) \nonumber \\
        &= \frac{1}{4|\mathscr{M}|^{2}} \int_{\mathscr{M}} d^{2}p d^{2}q \sum_{x_{1,2} \in \mathscr{O}} \langle x_{1}|\hat{A}|x_{2} \rangle e^{i(x_{2}(p + q) - x_{1}(p - q))} \nonumber \\
        &\times \sum_{y_{1,2} \in \mathscr{O}} \langle y_{1}|\hat{B}|y_{2} \rangle e^{i(y_{2}(p + q) - y_{1}(p - q))} \nonumber \\
        &\times \left( 4 + 2e^{\pm 2il_{1}q} + 2e^{\pm 2il_{2}q} + e^{\pm 2i(l_{1} + l_{2})q} + e^{\pm 2i(l_{1} - l_{2})q} \right) \nonumber \\
        &= \frac{1}{4} \sum_{x_{1,2} \in \mathscr{O}} \langle x_{1}|\hat{A}|x_{2} \rangle \sum_{y_{1,2} \in \mathscr{O}} \langle y_{1}|\hat{B}|y_{2} \rangle \big[ 4\delta_{2x_{1},2y_{2}}\delta_{2x_{2},2y_{1}} \nonumber \\
        &+ 2\delta_{2x_{1},2y_{2} \pm 2l_{1}}\delta_{2x_{2},2y_{1} \pm 2l_{1}} + 2\delta_{2x_{1},2y_{2} \pm 2l_{2}}\delta_{2x_{2},2y_{2} \pm 2l_{2}} \nonumber \\
        &+ \delta_{2x_{1},2y_{2} \pm 2(l_{1} + l_{2})}\delta_{2x_{2},2y_{1} \pm 2(l_{1} + l_{2})} \nonumber \\
        &+ \delta_{2x_{1},2y_{2} \pm 2(l_{1} - l_{2})}\delta_{2x_{2},2y_{1} \pm 2(l_{1} - l_{2})} \big] \nonumber \\
        &= \sum_{x_{1} \in \mathscr{O}} \langle x_{1}|\hat{A}\hat{B}|x_{1} \rangle = {\rm tr}\hat{A}\hat{B} \tag*{$\blacksquare$}
    \end{align*}
All Kronecker terms enclosed in square parentheses disappear for $x_{i},y_{i} \in \mathscr{O}$ save the first. Thus, a solution exists when $x_{1} = y_{2}$ and $x_{2} = y_{1}$. Following the example above, it can be simply demonstrated that:
    \begin{equation}
        {\rm Tr}_{\mathfrak{D}} \left( A_{W}B_{W} \right) = {\rm Tr}_{\mathfrak{D}} \left( A_{W} \star B_{W} \right) = {\rm tr}\hat{A}\hat{B}
    \end{equation}

\subsection{$W$-symbol of the identity operator}
\label{SectG10}

The identity operator's two-dimensional $W$-symbol is represented by:
    \begin{align}
        (\hat{1})_{W}(x,p)\Big|_{x \in \mathscr{D}} &= \frac{1}{4} \left[ 1 + \cos{2\pi c_{1}^{1}} + \cos{2\pi c_{2}^{1}} + \cos{2\pi \left( c_{1}^{1} + c_{2}^{1} \right)} \right] \nonumber \\
        &= 1
        \label{GJ04}
    \end{align}
where it is clear that for all values of $c_{1}^{1}$ and $c_{2}^{1}$, the identity operator is unitary. When validating (\ref{GE01}), consider the following:
    \begin{align*}
        &(\hat{1})_{W} = \int_{\mathfrak{M}} d^{2}q e^{2ixq}\langle p + q|p - q\rangle \nonumber \\
        &= \frac{1}{4} \int_{\mathfrak{M}} d^{2}q e^{2ixq} \delta^{\left[ \left( \Vec{g}_{1},\Vec{g}_{2} \right) \right]}(2q) f(q) \nonumber \\
        &= \frac{1}{16} \int_{\mathfrak{M}} d^{2}q e^{2ixq} \delta^{\left[ \left( \frac{1}{2}\Vec{g}_{1},\frac{1}{2}\Vec{g}_{2} \right) \right]}(q) \nonumber \\
        &\times \left( 1 + e^{-2il_{1}q} + e^{-2il_{2}q} + e^{-2i(l_{1} + l_{2})q} \right) \nonumber \\
        &= \frac{1}{16} \int_{-\frac{2\pi}{3\ell}}^{\frac{2\pi}{3\ell}} dq_{x} e^{2ixq_{x}} \int_{-\frac{\pi}{\sqrt{3}\ell}}^{\frac{\pi}{\sqrt{3}\ell}} dq_{y} e^{2iyq_{y}} \bigg( \delta(q_{x})\delta(q_{y}) \nonumber \\
        &+ \delta \left( q_{x} \mp \frac{\pi}{3\ell} \right) \delta \left( q_{y} \pm \frac{\pi}{\sqrt{3}\ell} \right) \nonumber \\
        &+ \delta \left( q_{x} \pm \frac{\pi}{3\ell} \right) \delta \left( q_{y} \pm \frac{\pi}{\sqrt{3}\ell} \right) + \delta \left( q_{x} \pm \frac{2\pi}{3\ell} \right) \delta(q_{y}) \bigg) \nonumber \\
        &\times \bigg( 1 + 2e^{-i3\ell q_{x}}\cos{\sqrt{3}\ell q_{y}} + e^{-i6\ell q_{x}} \bigg) \nonumber \\
        &= \frac{1}{4} + \frac{1}{8}e^{i\frac{2\pi}{3\ell}x}e^{-i\frac{2\pi}{\sqrt{3}\ell}y} + \frac{1}{8}e^{-i\frac{2\pi}{3\ell}x}e^{i\frac{2\pi}{\sqrt{3}\ell}y} \nonumber \\
        &+ \frac{1}{8}e^{-i\frac{2\pi}{3\ell}x}e^{-i\frac{2\pi}{\sqrt{3}\ell}y} + \frac{1}{8}e^{i\frac{2\pi}{3\ell}x}e^{i\frac{2\pi}{\sqrt{3}\ell}y} + \frac{1}{8}e^{-i\frac{4\pi}{3\ell}x} + \frac{1}{8}e^{i\frac{4\pi}{3\ell}x} \nonumber \\
        &= \frac{1}{4} + \frac{1}{2}\cos{\frac{2\pi}{3\ell}x}\cos{\frac{2\pi}{\sqrt{3}\ell}y} + \frac{1}{4}\cos{\frac{4\pi}{3\ell}x},
    \end{align*}
where
    \begin{gather*}
        x = \left( c_{1}^{1}\Vec{l}_{1} + c_{2}^{1}\Vec{l}_{2} \right)_{x} = \frac{3\ell}{2} \left( c_{1}^{1} + c_{2}^{1} \right) \nonumber \\
        y = \left( c_{1}^{1}\Vec{l}_{1} + c_{2}^{1}\Vec{l}_{2} \right)_{y} = \frac{\sqrt{3}\ell}{2} \left( c_{1}^{1} - c_{2}^{1} \right) \tag*{$\blacksquare$}
    \end{gather*}
These variables, $x$ and $y$, are the linear projections of the extended lattice vectors (lattice $\mathfrak{D}$) onto the axes.

\subsection{Star product without differentiation}

\label{SectWStar}
One can represent the star product of Weyl symbols for $x \in \mathscr{S}$ through the matrix elements of $\hat A$ and $\hat B$. Using Eq. (\ref{GWxAH1}) we obtain (for $x \in \mathscr{S}$):
\begin{eqnarray}
	&&A_{W}(x,p) \star B_{W}(x,p)
	=  \sum_{2z,2\bar{z},u,\bar{u} \in \mathscr{D}} \delta_{2x-\bar{u},2z} \delta_{2x+{u},2\bar{z}}\nonumber\\&&\int \frac{\D{p}^\prime}{|{\mathfrak{M}}|}\frac{d\bar{p}^\prime}{|{\mathfrak{M}}|}e^{ ip^\prime u + i \bar{p}^\prime \bar{u}} A_{W}(z,p-p^\prime)B_{W}(\bar{z},p-\bar{p}^\prime)\nonumber\\&=&
	\sum_{z,\bar{z} \in \mathscr{S}} \int \frac{\D{p}^\prime}{|{\mathfrak{M}}|}\frac{d\bar{p}^\prime}{|{\mathfrak{M}}|}\nonumber\\&&e^{ 2ip^\prime (\bar{z}-x) + i \bar{p}^\prime (x-z)} A_{W}(z,p-p^\prime)B_{W}(\bar{z},p-\bar{p}^\prime)
\end{eqnarray}
One can see, that in order to define the star product of the  symbols $A_{W}(x,p)$ and $B_{W}(x,p)$ for $x \in \mathscr{S}$ we do not need to know the values of these functions for all real values of $x$. It is enough to know the values of the Weyl symbols for $x \in \mathscr{S}$. Moreover, according to Eq. (\ref{BPS0}) we have
\begin{equation}
	{A}_{W}(x,p)\Big|_{x
		\in \mathscr{S}\setminus \mathscr{D}}=0\label{BPS01}
\end{equation}
for the operators that obey  Eq. (\ref{GF03}). As a result for such operators all needed information is encoded in the Weyl symbols $A_W(x,p)$ and $B_W(x,p)$ for $x \in \mathscr{D}$:
\begin{eqnarray}
	&&A_{W}(x,p) \star B_{W}(x,p)\Big|_{x \in \mathscr{D}}
	=
	\sum_{z,\bar{z} \in \mathscr{D}} \int \frac{\D{p}^\prime}{|{\mathfrak{M}}|}\frac{d\bar{p}^\prime}{|{\mathfrak{M}}|}\nonumber\\&&e^{ 2ip^\prime (\bar{z}-x) + 2i \bar{p}^\prime (x-z)} A_{W}(z,p-p^\prime)B_{W}(\bar{z},p-\bar{p}^\prime)
\end{eqnarray}



\section{Dynamics of systems defined on finite lattice, and Hall conductivity }
\label{Dynamics}

\subsection{Keldysh technique of field theory }

\label{SectKeldysh}


{Here we follow the approach developed in  \cite{Sugimoto2008} and \cite{BFLZZ2021}. We work in the relativistic system of units with $\hbar = 1$. Moreover, we absorb electric charge in the definition of electromagnetic field. In order to come back to the usual system of units expression for the conductivity to be obtained below should be multiplied by $e^2/\hbar$. 
	
Let us consider the system defined on the honeycomb lattice $\mathcal O$. Time is not discretized and is continuous. Field theory Hamiltonian is denoted by    $\hat\cH$.  An operator  $O[\psi,\bar{\psi}]$ is a functional of fields $\hat{\psi}, \hat{\bar{\psi}}$. Operator $O$ at time $t$ is a function of $\psi$ and $\bar{\psi}$  defined at the same time. Average of the corresponding quantity is given by
	$$
	\langle O \rangle
	=  {\tr} \,\Bigl(\hat{\rho}(t_i)\, e^{- i \int_{t_i}^{t} \hat\cH dt }  O[\hat{\psi},\hat{\bar{\psi}}] e^{- i \int_{t}^{t_f} \hat\cH dt } e^{ i \int_{t_i}^{t_f} \hat\cH dt }\Bigr).
	$$
	Here $t_i < t < t_f$, and  $\hat{\rho}(t_i)$ is density matrix at $t_i$. Time ordering $T$ allows us to rewrite the above expression as
	$$
	\langle O \rangle =  {\tr} \,\Bigl(T\,\Big[\hat{\rho}(t_i)\, e^{- i \int_{t_i}^{t_f} \hat\cH dt }  O[\hat{\psi},\hat{\bar{\psi}}]\Big] e^{ i \int_{t_i}^{t_f} \hat\cH dt }\Bigr).
	$$
	For the considered lattice system  $\langle O \rangle$ is given by
	$$
	\langle O \rangle
	=  \int {\cal D}\bar{\psi} {\cal D} \psi\, O[\psi,\bar{\psi}]
	\exp\left\{\ii \int_C dt \sum_x \, \bar{\psi}(t,x) \hat{Q} \psi(t,x) \right\}.
	$$
	Here $\psi$ and $\bar{\psi}$ are the Grassmann variables,  $x$ is a  lattice point. Without interactions $\hat{Q}$ is  $\hat{Q} = i \partial_t-\hat{H}$, where $\hat{H}$ is the one-particle Hamiltonian.
	Integration over time is along the Keldysh contour $C$. The contour starts at the $t_i$, goes to  $t_f$, and returns back from $t_f$ to $t_i$.
	
	The forward part of the contour carries fields $\bar{\psi}_-(t,x)$ and $\psi_-(t,x)$. The fields on the  backward part are  $\bar{\psi}_+(t,x)$ and $\psi_+(t,x)$.
	
	The boundary conditions relate fields of the forward and backward parts of the Keldysh contour: $\bar{\psi}_-(t_f,x) =  \bar{\psi}_+(t_f,x)$ and $\psi_-(t_f,x)=\psi_+(t_f,x)$. The  integration measure ${\cal D} \bar{\psi} {\cal D} \psi$ contains  $\bar{\psi}_+(t_i,x)$, ${\psi}_+(t_i,x)$ and $\bar{\psi}_-(t_i,x)$, ${\psi}_-(t_i,x)$ and depends on initial density matrix $\hat{\rho}$:
	\begin{widetext}
	\begin{eqnarray}
		\langle O \rangle
		&=&
		\int \frac{{\cal D}\bar{\psi}_\pm {\cal D} \psi_\pm}{{\rm Det}\, (1+\rho)} \, O[\psi_+,\bar{\psi}_+]\nonumber\\
		&&
		\qquad	{\rm exp}\left\{\ii \int_{t_i}^{t_f} dt \sum_x \[\bar{\psi}_-(t,x) \hat{Q} \psi_-(t,x)-\bar{\psi}_+(t,x) \hat{Q} \psi_+(t,x)\]-\sum_x\, \bar{\psi}_-(t_i,x) {\rho} \psi_+(t_i,x)\right\} .\label{eq1}
	\end{eqnarray}
\end{widetext}
 $\rho$ is an operator in one - particle Hilbert space. Its eigenstates are $|\lambda_i \rangle$, the matrix elements enter expression
	$\frac{\langle \lambda_i |\rho|\lambda_i\rangle}{1+\langle \lambda_i |\rho|\lambda_i\rangle}$ for  the probability that the one - particle state $|\lambda_i\rangle$ is occupied. At the same time  $\frac{1}{1+\langle \lambda_i |\rho|\lambda_i\rangle}$ is the probability that the same state is empty.
	Keldysh spinors are composed of
	\be
	\Psi = \left(\begin{array}{c}\psi_-\\ \psi_+ \end{array}\right),
	\label{KelPsi}
	\ee
	The average of an operator $O$ is
	\begin{widetext}	
	\begin{eqnarray}
		\langle O \rangle
		&=& \frac{1}{{\rm Det}\, (1+\rho)}\int {\cal D}\bar{\Psi} {\cal D} \Psi \,
		O[\Psi,\bar\Psi]\,
		{\rm exp}\Bigl\{\ii \int_{t_i}^{t_f} dt \sum_x \bar{\Psi}(t,x) \hat{\bf Q} \Psi(t,x) \Bigr\} .
	\end{eqnarray}
\end{widetext}
	
	\rv{Here $x$ is a two - dimensional vector.}  $\bm {\hat Q}$ is given in Keldysh representation as
	\begin{eqnarray}
		\hat{\bf Q}
		= \left(\begin{array}{cc}Q_{--} & Q_{-+}\\ Q_{+-} & Q_{++} \end{array} \right).
		\label{KelQ}
	\end{eqnarray}
To calculate components of this matrix one should use continuum limit of lattice regularized expressions:
	\begin{eqnarray}
		Q_{++}  &=& -\Big(\ii \partial_t-\hat{H} - \ii \epsilon \frac{1-\rho}{1+\rho}\Big), \nonumber \\
		Q_{--}  &=&  \ii \partial_t-\hat{H} + \ii \epsilon \frac{1-\rho}{1+\rho}, \nonumber \\
		Q_{+-}  &=&  -2\ii \epsilon \frac{1}{1+\rho}, \nonumber \\
		Q_{-+}  &=& 2\ii  \epsilon \frac{\rho}{1+\rho}
		\label{Qnaive} .
	\end{eqnarray}
	Here $\rho$ gives rise to one - particle distribution $f = \rho (1+\rho)^{-1}$. For the distribution depending only on energy  $\rho = \rho(\hat{H})$ is a function of the one - particle Hamiltonian. The infinitely small parameter $\epsilon \to 0$ points out the way to avoid the singularities while calculating the inverse operators (see Sect. 5.1 of \cite{Kamenev2}).

The Green function $\hat{\bf G}$ is defined as
\begin{widetext}
\begin{eqnarray}
	G_{\alpha_1 \alpha_2}(t,x|t^\prime,x^\prime)
	&=& \int \frac{{\cal D}\bar{\Psi} {\cal D} \Psi}{\ii{\rm Det}\, (1+\rho)}
	\Psi_{\alpha_1}(t,x) \bar{\Psi}_{\alpha_2}(t^\prime,x^\prime)
	\, \exp\left\{\ii \int_{t_i}^{t_f} dt \sum_x\, \bar{\Psi}(t,x) \hat{\bf Q} \Psi(t,x) \right\}.\label{G1}
\end{eqnarray}
\end{widetext}
Here $\alpha$ is Keldysh spinor index  \Ref{KelPsi}. We have an equation
$$
\hat{\bf Q}\hat{\bf G}=1 .
$$
Components of $\hat{\bf G}$ obey
\be
G_{--}+G_{++}-G_{-+}-G_{+-}=0
\label{G-rel}
\ee
while 	
\be	
Q_{--}+Q_{++}+Q_{-+}+Q_{+-}=0 .
\label{Q-rel}
\ee

Another representation is related to the spinors defined above as	
$$
\begin{pmatrix}\psi_1 \\ \psi_2 \end{pmatrix}=\frac{1}{\sqrt{2}}\begin{pmatrix}1 & 1 \\ 1 & -1 \end{pmatrix}\begin{pmatrix}\psi_- \\ \psi_+ \end{pmatrix}
$$$$
\begin{pmatrix}\bar{\psi}_1 & \bar{\psi}_2 \end{pmatrix}=\frac{1}{\sqrt{2}}\begin{pmatrix}\bar{\psi}_- & \bar{\psi}_+ \end{pmatrix}\begin{pmatrix}1 & 1 \\ -1 & 1 \end{pmatrix}.
$$
Green function in this representation is triangular
\begin{eqnarray}
	\hat{\bf G}^{(K)}&=&-i\langle \begin{pmatrix}\psi_1 \\ \psi_2 \end{pmatrix}\otimes\begin{pmatrix}\bar{\psi}_1 & \bar{\psi}_2 \end{pmatrix}\rangle \\\nonumber&=&\frac{1}{2}\begin{pmatrix}1 & 1 \\ 1 & -1 \end{pmatrix}\begin{pmatrix}G^{--} & G^{-+} \\ G^{+-} & G^{++} \end{pmatrix}\begin{pmatrix}1 & 1 \\ -1 & 1 \end{pmatrix}\\\nonumber
	&=&\begin{pmatrix}
		G^\rR &G^\rK \\0&G^\rA
	\end{pmatrix}.
	\label{GK}
\end{eqnarray}
We introduced above Keldysh, Advanced and Retarded Green  functions:
\bes
G^\rK  & =G^{-+}+G^{+-}=G^{--}+G^{++},\\
G^\rA  & =G^{--}-G^{+-}=G^{-+}-G^{++},\\
G^\rR  & =G^{--}-G^{-+}=G^{+-}-G^{++}.
\end{eqsplit}

Another triangular representation will be used below
\bes
\hat{\bf G}^{(<)}&=\begin{pmatrix}
1&1\\0&1
\end{pmatrix}\begin{pmatrix}
G^\rR &G^\rK \\0&G^\rA
\end{pmatrix}\begin{pmatrix}
1&-1\\0&1
\end{pmatrix}
\\
&=	\begin{pmatrix}
G^\rR &2G^<\\0&G^\rA
\end{pmatrix} .
\label{G<}	
\end{eqsplit}
It is expressed through the Green function defined by Eq. (\ref{G1}) as
\be
\hat{\bf G}^{(<)} = U \hat{\bf G} V,
\ee
where
$$
U=\frac{1}{\sqrt{2}}\begin{pmatrix}
1&1\\0&1
\end{pmatrix}\begin{pmatrix}1 & 1 \\ 1 & -1 \end{pmatrix}=\frac{1}{\sqrt{2}}\begin{pmatrix}2 & 0\\1 &-1 \end{pmatrix}
$$
and
$$
V=\frac{1}{\sqrt{2}}\begin{pmatrix}
1&1\\-1&1
\end{pmatrix}\begin{pmatrix}1 & -1 \\ 0 & 1 \end{pmatrix}=\frac{1}{\sqrt{2}}\begin{pmatrix}1 & 0\\-1& 2 \end{pmatrix}.
$$
In addition, we have
\begin{eqnarray}
\hat{\bf Q}^{(<)}&=&V^{-1}\hat{\bf Q} U^{-1}\\\nonumber&=&\frac{1}{2}\begin{pmatrix}2 & 0\\1& 1 \end{pmatrix}\begin{pmatrix}Q^{--} & Q^{-+} \\ Q^{+-} & Q^{++} \end{pmatrix}\begin{pmatrix}1 & 0\\1& -2 \end{pmatrix}\\\nonumber&=&
\begin{pmatrix}Q^{--}+Q^{-+} & -2Q^{-+} \\ \frac{Q^{--}+Q^{+-}+Q^{-+}+Q^{++}}{2} & -Q^{-+}-Q^{++} \end{pmatrix}\\\nonumber&=&\begin{pmatrix}Q^\rR  & 2Q^< \\ 0 & Q^\rA  \end{pmatrix},
\end{eqnarray}
Here we denote
\be
Q^\rR =Q^{--}+Q^{-+}, \qquad
Q^\rA =-Q^{-+}-Q^{++}, \qquad
Q^<=-Q^{-+},
\label{Qar_def}
\ee
Then
\begin{equation}
G^\rA  = (Q^\rA )^{-1}, \qquad G^\rR  = (Q^\rA )^{-1}, \qquad G^< =-G^\rR  Q^< G^\rA .
\label{Gar_def}
\end{equation}
with
\bes
G^\rR
&= (\ii \partial_t-\hat{H}e^{+ \epsilon \partial_t})^{-1}
= (\ii \partial_t-\hat{H}+\ii \epsilon )^{-1},
\\
G^\rA  &= ( \ii \partial_t-\hat{H}e^{- \epsilon \partial_t})^{-1}
= ( \ii \partial_t-\hat{H}-\ii \epsilon )^{-1},
\\
G^< &=(G^\rA -G^\rR ) \frac{\rho}{\rho+1}.
\label{Gar_expl}
\end{eqsplit}
 $\hat{\bf Q}^<$  is inverse to $\hat{\bf G}^<$:
\bes
Q^{<}&=(Q^\rA -Q^\rR )\frac{\rho}{\rho+1} = -2\ii\epsilon \frac{\rho}{\rho+1},
\\
Q^\rR  &= \ii \partial_t-\hat{H}+\ii \epsilon ,	
\\
Q^\rA  &=  \ii \partial_t-\hat{H}-\ii \epsilon .
\label{Qar_expl}
\end{eqsplit}
(See  \cite{Kamenev,Kamenev2}.)

\subsection{ Electric conductivity and Wigner - Weyl calculus}

Here we adopt basic notions of Wigner - Weyl calculus  \cite{ZW2019,Sugimoto} to the models defined on honeycomb lattice. The $2+1$ dimensional vectors are denoted by large Latin letters. For any operator $\hat{A}$ its matrix elements in momentum space are denoted by $A(P_1,P_2) = \langle P_1 | \hat{A} | P_2 \rangle $.  Correspondingly, the space components of momentum belong to the Brillouin zone while its time component (frequency) is real - valued. We then define Weyl symbol of an operator $\hat A$ as the mixture of lattice Weyl symbol and Wigner transformation with respect to the frequency component:
\begin{widetext}
	\begin{eqnarray}
		A_W(X|P)&=&2\int d P^0\,\int_{\mathscr{M}} d^{2}\vec{Q}\,  e^{-2 \ii X^\mu Q_\mu }  A(P+Q,P-Q) \left( 1 + e^{-2il_{1}\vec{Q}} + e^{-2il_{2}\vec{Q}} + e^{-2i(l_{1} + l_{2})\vec{Q}} \right) \label{WignerTr} \\ && \quad \mu =0,1,2\nonumber
	\end{eqnarray}
\end{widetext}
$2+1$ D momentum is $P^\mu=(P^0,p)$, and $P_\mu = (P^0,-p)$.
Weyl symbol of Keldysh Green function $\hat{\bf G}$ is denoted by $\hat{G}$, while Weyl symbol of Keldysh operator $\hat{\bf Q}$ is $\hat{Q}$. The subscript $W$ is omitted below for brevity.

 $\hat{G}$ and $\hat{Q}$ obey Groenewold equation
\begin{equation}
\hat{Q} * \hat{G} = 1_W.
\end{equation}
Moyal product $*$ is written as
\begin{equation}
\left(A* B\right)(X|P) = A(X|P)\,e^{\rv{-}\ii(\overleftarrow{\partial}_{X^{\mu}}\overrightarrow{\partial}_{P_{\mu}}-\overleftarrow{\partial}_{P_{\mu}}\overrightarrow{\partial}_{X^{\mu}})/2}B(X|P).
\end{equation}
Electromagnetic potential $A$ corresponds to constant components of electric field with the field strength ${\cal F}^{\mu\nu}$. We choose the gauge, in which the spatial part of potential is proportional to time but does not depend on spatial coordinates.

\subsection{Gauge transformation of Weyl symbol}

\label{AWS}

Weyl symbol of operator $\hat A$ may be represented also as
\begin{widetext}
	\begin{eqnarray}
		A_W(X|P)&=& 2\int d Z^0 d Y^0\, \sum_{\vec{Z},\vec{Y}\in \mathscr{O}; n_i = 0,1} \,  e^{ \ii (Z^\mu-Y^\mu) P_\mu }  \bra{Z} \hat{A} \ket{Y} \delta(2 X^0- Z^0-Y^0){\bf d}(2\vec{X} - \vec{Z}-\vec{Y} - 2 l_i n_i) \label{WignerTr2}
	\end{eqnarray}
\end{widetext}

$U(1)$ gauge transformation acts as
$\ket{X} \to e^{i \alpha(X)} \ket{X}$. As a result Weyl symbol of an operator $\hat{A}$ is transformed as
\begin{widetext}
	\begin{eqnarray}
		A_W(X|P)&\to& 2\int d Z^0 d Y^0\, \sum_{\vec{Z},\vec{Y}\in \mathscr{O}; n_i = 0,1} \,  e^{ \ii (Z^\mu-Y^\mu) P_\mu + i (\alpha(Y)-\alpha(Z)) }  \bra{Z} \hat{A} \ket{Y} \delta(2 X^0- Z^0-Y^0){\bf d}(2\vec{X} - \vec{Z}-\vec{Y} - 2 l_i n_i) \nonumber
	\end{eqnarray}
\end{widetext}
Let us consider those gauge transformations, for  which function $\alpha$ almost does not vary at the distances of the order of the correlation length $\lambda$ characterizing operator $\hat{A}$, i.e. $|\lambda \partial \alpha|\ll 1$. We call these transformations "slow" (with respect to $\hat A$). For them we obtain:
\begin{widetext}
	\begin{eqnarray}
		A_W(X|P)&\to& 2\int d Z^0 d Y^0\, \sum_{\vec{Z},\vec{Y}\in \mathscr{O}; n_i = 0,1} \,  e^{ \ii (Z^\mu-Y^\mu) (P_\mu - \partial_\mu \alpha(X)) }  \bra{Z} \hat{A} \ket{Y} \delta(2 X^0- Z^0-Y^0){\bf d}(2\vec{X} - \vec{Z}-\vec{Y} - 2 l_i n_i) \nonumber\\&=&	A_W(X|P-\partial_\mu \alpha(X))
	\end{eqnarray}
\end{widetext}
If operator $\hat A$ depends on the $U(1)$ gauge field $A$ then we may require that the gauge transformation of $\hat A$ should be compensated by the gauge transformation of field $A$. This occurs, for example, for Dirac operator $\hat{Q}$ due to gauge invariance of the whole model. Consideration of "slow" gauge transformation results in the requirement that Weyl symbol $A_W(x,p)$ depends on $A(x)$ through the functional dependence on $P - A(x)$, and gauge invariant quantities: field strength $F_{ij}$ and its derivatives, provided that variation of $A(x)$ may be neglected at the distances of the order of $\lambda$, i.e. $|\lambda^2 F_{ij}| \ll 1$. As a result for such $A(x)$ we may represent $A_W$ as a series
\begin{eqnarray}
	A_W(X|P)&=&A^{(0)}_W(X|P - A(x))\nonumber\\&&+B^{(1)}_{(ij) W}(X|P - A(x))F_{ij}(X)\nonumber\\&&+A^{(2)}_{(ijk) W}(X|P - A(x))\partial_kF_{ij}(x) + ...\label{BWA}
\end{eqnarray}
Here dots denote the higher  order terms in derivatives.
This expansion is reasonable, i.e. the higher order terms are smaller than the lower order terms under the same condition $|\lambda^2 F_{ij}| \ll 1$.

In particular, for bare $\hat Q$ the correlation length $\lambda$ is given by the lattice spacing, and we may use the above expansion for the fields $A$ that vary slowly at the distance of the order of lattice spacing. Response to such fields gives electric conductivity.


Thus in order to calculate conductivity we  expand our expressions in powers of
${\cal F}^{\mu \nu}$ up to the linear term. We denote  $\pi = P- A$. Here $\pi^\mu$ is $2+1$ - dimensional vector similar to $P^\mu$. The Moyal product may be decomposed as
\begin{equation}
* =  \star~ e^{\rv{-}\ii  \mathcal{F}^{\mu\nu}\overleftarrow{\partial}_{\pi^{\mu}}\overrightarrow{\partial}_{\pi^{\nu}}/2}.
\end{equation}
with
\begin{equation}
\left(A\star B\right)(X|\pi) = A(X|\pi)\,e^{\rv{-\ii(\overleftarrow{\partial}_{X^{\mu}}\overrightarrow{\partial}_{\pi_{\mu}}-\overleftarrow{\partial}_{\pi_{\mu}}\overrightarrow{\partial}_{X^{\mu}})/2}}B(X|\pi).
\end{equation}
If external field $A$ does not depend on spatial coordinates, then this expression can be used.

Both $\hat{Q}$ and $\hat{G}$ are expanded in powers of  $\mathcal{F}^{\mu\nu}$ up to the linear terms
\begin{equation}
\hat{Q} = \hat{Q}^{(0)}  +\frac{1}{2}\mathcal{F}^{\mu\nu}\hat{Q}_{\mu\nu}^{(1)},\quad
\hat{G} = \hat{G}^{(0)}  +\frac{1}{2}\mathcal{F}^{\mu\nu}\hat{G}_{\mu\nu}^{(1)}.\label{QGK}
\end{equation}
We omit below the superscript $^{(0)}$ for brevity. The Green function (and its inverse) are expressed through the one - particle Hamiltonian as
\bes
G^\rR
&=(\pi_0-\hat{H}(\vec{\pi},x)+\ii \epsilon )^{-1},
\\
G^\rA  &= ( \pi_0-\hat{H}(\vec{\pi},x)-\ii \epsilon )^{-1},
\\
G^< &=(G^\rA -G^\rR ) f(\pi_0) = 2\pi i \delta(\pi_0-\hat{H}(\vec{\pi}))f(\pi_0).
\label{Gar_expl}
\end{eqsplit}
 $\hat{\bf Q}^<$ is inverse to $\hat{\bf G}^<$:
\bes
Q^{<}&=(Q^\rA -Q^\rR )f(\pi_0) = -2\ii\epsilon f(\pi_0),
\\
Q^\rR  &= \pi_0-\hat{H}(\vec{\pi},x)+\ii \epsilon ,	
\\
Q^\rA  &=  \pi_0-\hat{H}(\vec{\pi},x)-\ii \epsilon .
\end{eqsplit}
 Groenewold equation can be written as
\begin{equation}
\left(\hat{Q}  +\frac{1}{2}\mathcal{F}^{\mu\nu}\hat{Q}_{\mu\nu}^{(1)}\right)\star~ e^{\rv{-}i  \mathcal{F}^{\mu\nu}\overleftarrow{\partial}_{\pi^{\mu}}\overrightarrow{\partial}_{\pi^{\nu}}/2}\left(\hat{G}  +\frac{1}{2}\mathcal{F}^{\mu\nu}\hat{G}_{\mu\nu}^{(1)}\right) = 1_W.
\label{Groe-F}
\end{equation}
In the zeroth order in $\cal F$ it is reduced to $\hat{Q} \star \hat{G}  = 1_W$ (in the following we will write $\hat{G}$ instead of $\hat{G}^{(0)}$ if this will not lead to contradictions), while the first order gives
$\hat{Q} \star \hat{G}^{(1)}+\hat{Q}^{(1)}\star\hat{G} \rv{-} \ii \hat{Q} \star \overleftarrow{\partial}_{\pi^{\mu}}\overrightarrow{\partial}_{\pi^{\nu}} \hat{G}  = 0$.

At this point we notice that according to the properties of the Weyl symbols $Q_W(x,p)$ and $G(x,p)$, all needed information is encoded in their values at $x \in \mathscr{D}$. At these values $1_W(x,p) \equiv 1$. As a result  $\hat{G}^{(0)}(x,p)$ is smooth function of $x$ and $p$ as long as  $\hat{Q}^{(0)} \equiv  \hat{Q}^{}$ is smooth function of $x$ and $p$. Therefore, the expansion in powers of derivatives has sense. If we would use instead of the Weyl symbol  $\hat{G}_W$ the $\mathscr{B}$ symbol  $\hat{G}_{\mathscr{B}}$, then function  $\hat{G}^{(0)}_{\mathscr{B}}(x,p)$ will be oscillating fast since $1_{\mathscr{B}}(x,p)$ is fast oscillating. The derivative expansion of such functions is problematic, and most likely we cannot expand our expressions in powers of $F_{ij}$ (although this question is to be analyzed more carefully).

We obtain
\begin{eqnarray}
&&\hat{G}_{\mu\nu}^{(1)} =-\hat{G} \star  \hat{Q}_{\mu\nu}^{(1)}\star \hat{G} \nonumber\\ && \rv{-}   \ii\left(\hat{G} \star \partial_{\pi^{\mu}}\hat{Q}  \star\hat{G} \star \partial_{\pi^{\nu}}\hat{Q} \star \hat{G} -(\mu\leftrightarrow \nu)\right)/{2}
.
\label{QGK1}
\end{eqnarray}

The above derivation is somehow similar to the derivation presented in \cite{Sugimoto}. The difference is that here we consider the lattice model. For the homogeneous system Weyl symbol $\hat{Q}$ does not depend on coordinate $x$. It may depend on $P^0$ and $X^0$ and on $p \in {\cal M}$. Electric current density is given by
$$
\hat{j^i}=-\hat{\bar{\psi}} \frac{\partial \hat{Q}}{\partial p_i} \hat{\psi},\quad
i=1,2,\ldots D.
$$
Recall that spatial components of momentum are $p^i=p_i = P^i = - P_i$.
\begin{eqnarray}
\langle j^i(t,x) \rangle
&=&-\frac{\ii}{2}{\tr} \[\hat{\bf G} \hat{{\bf v}}^i\].
\end{eqnarray}
Velocity operator is
$$
\hat{\bf v}^i = \partial_{p_i}\begin{pmatrix}-Q^{--} & 0 \\ 0 & Q^{++} \end{pmatrix}.
$$

For the non - uniform systems the above  expression for the electric current density is not valid. Nevertheless, response of the partition function to variation of electromagnetic potential gives expression for the electric current averaged over the whole area:
\begin{eqnarray}
\langle J^i(t) \rangle
&=&-\frac{\ii}{2} \frac{1}{|\mathfrak{D}|} \int \frac{dP^0}{2\pi} \int_{\mathscr{M}} \frac{d^{2}\vec{P}}{(2\pi)^2}\, \sum_{x\in \mathfrak{D}} {\tr}\,  {\bf G}(X|P) \nonumber\\&&\partial_i \begin{pmatrix}-Q^{--}(X|P) & 0 \\ 0 & Q^{++}(X|P) \end{pmatrix}
\end{eqnarray}
Here we use  Weyl symbols of operators.

$$
\hat{\bf v}^i = \partial_{p_i}\begin{pmatrix}-Q^{--}(P|X) & 0 \\ 0 & Q^{++}(P|X) \end{pmatrix}.
$$
is Weyl symbol of velocity operator.

Average current $J$ may be expressed  through the Keldysh Green function  in the representation of Eq. (\ref{G<}).
\begin{eqnarray}
\hat{\bf v}_i^{(<)}&=&\partial_{p_i}\frac{1}{2}\begin{pmatrix}2 & 0\\1& 1 \end{pmatrix}\begin{pmatrix}-Q^{--} & 0 \\ 0 & Q^{++} \end{pmatrix}\begin{pmatrix}1 & 0\\1& -2 \end{pmatrix}\\\nonumber
&=&\partial_{p_i}\begin{pmatrix} -Q^{--} & 0 \\ \frac{-Q^{--}+Q^{++}}{2}  & -Q^{++} \end{pmatrix}
\end{eqnarray}
Using Eq. \Ref{Qar_def} ($Q^{--}=Q^\rR +Q^<$, $Q^{-+}=-Q^<$, $Q^{+-}=-Q^\rR +Q^\rA -Q^<$, and $Q^{++}=Q^<-Q^\rA $) we represent current density as
\begin{eqnarray}
&&\langle J^i \rangle
=- \frac{\ii}{2 |\mathscr{O}|}  \int \frac{dP^0}{2\pi}  \Tr\left[\hat{\bf G}\hat{\bf v}^i\right]\nonumber\\&
=&-\frac{\ii}{2 |\mathscr{O}|} \int \frac{dP^0}{2\pi} \Tr\Bigl[\begin{pmatrix}
G^\rR &2G^<\\0&G^\rA
\end{pmatrix}\nonumber\\&&\partial_{p_i}\begin{pmatrix}- Q^\rR -Q^< & 0 \\ -\frac{Q^\rR +Q^\rA }{2}  & -Q^<+Q^\rA  \end{pmatrix}\Bigr]\\\nonumber
&=&\frac{\ii}{2 |\mathscr{O}|} \int \frac{dP^0}{2\pi} \Tr \left(G^\rR \partial_{p_i} Q^\rR -G^\rA \partial_{p_i}Q^\rA \right)
\nonumber\\&&+\frac{\ii}{2 |\mathscr{O}|} \int \frac{dP^0}{2\pi} \Tr \left(G^\rR \partial_{p_i} Q^<+G^<\partial_{p_i} Q^\rA \right)\nonumber\\&&
+\frac{\ii}{2 |\mathscr{O}|} \int \frac{dP^0}{2\pi} \Tr \left(G^\rA \partial_{p_i}Q^< +G^<\partial_{p_i} Q^\rR \right)
\end{eqnarray}
The second term here is expressed through $\frac{\ii}{2}\Tr\left({G}\partial_{p_i} {Q}\right)^<$.
 We get
\begin{eqnarray}
\langle J^i \rangle
&=&\frac{\ii}{2 |\mathscr{O}|} \int \frac{dP^0}{2\pi}  \Tr\left(\hat{\bf G}\partial_{p_i}\hat {\bf Q}\right)^\rR
\nonumber\\&&+ \frac{\ii}{2 |\mathscr{O}|} \int \frac{dP^0}{2\pi}  \Tr\left(\hat{\bf G}\partial_{p_i}\hat {\bf Q}\right)^<+{\rm c.c.}
\label{<j>}
\end{eqnarray}
Electric current is given by
\begin{widetext}
\bes
\langle J^i(t) \rangle
& 	= \rv{-} \frac{\ii }{2} \frac{1}{|\mathfrak{D}|} \int \frac{dP^0}{2\pi} \int_{\mathscr{M}} \frac{d^{2}\vec{P}}{(2\pi)^2}\, \sum_{x\in \mathfrak{D}}
\tr\left(\hat{G} (\partial_{\pi_{i}}\hat{Q})\right)^{\rR}
\rv{-} \frac{\ii }{2}
\frac{1}{|\mathfrak{D}|} \int \frac{dP^0}{2\pi} \int_{\mathscr{M}} \frac{d^{2}\vec{P}}{(2\pi)^2}\, \sum_{x\in \mathfrak{D}}
\tr\left(\hat{G} (\partial_{\pi_{i}}\hat{Q})\right)^{\rA}	
\\
&\qquad		
\rv{-}\frac{\ii }{2}
\frac{1}{|\mathfrak{D}|} \int \frac{dP^0}{2\pi} \int_{\mathscr{M}} \frac{d^{2}\vec{P}}{(2\pi)^2}\, \sum_{x\in \mathfrak{D}}
\tr\left(\hat{G} (\partial_{\pi_{i}}\hat{Q})\right)^{<}
\rv{-}\frac{\ii }{2}
\frac{1}{|\mathfrak{D}|} \int \frac{dP^0}{2\pi} \int_{\mathscr{M}} \frac{d^{2}\vec{P}}{(2\pi)^2}\, \sum_{x\in \mathfrak{D}}
\tr\left((\partial_{\pi_{i}}\hat{Q}) \hat{G}\right)^{<} .
\end{eqsplit}
\end{widetext}
The poles of $G^{\rR}$  ($G^{\rA}$) are shifted out of the real axis of frequency $\omega$. The integration is closed at infinity if we use lattice regularization of time. As a result  the sum of the first two terms vanishes in the above expression:
\begin{widetext}
\be
J^i(t) = \rv{-}\frac{\ii }{2}\frac{1}{|\mathfrak{D}|} \int \frac{dP^0}{2\pi} \int_{\mathscr{M}} \frac{d^{2}\vec{P}}{(2\pi)^2}\, \sum_{x\in \mathfrak{D}}
\tr\left(\hat{G} (\partial_{\pi_{i}}\hat{Q})\right)^{<}
\rv{-}\frac{\ii }{2}
\frac{1}{|\mathfrak{D}|} \int \frac{dP^0}{2\pi} \int_{\mathscr{M}} \frac{d^{2}\vec{P}}{(2\pi)^2}\, \sum_{x\in \mathfrak{D}}
\tr\left((\partial_{\pi_{i}}\hat{Q}) \hat{G}\right)^{<} .
\label{J Wigner}
\ee
\end{widetext}
Using Eqs. (\ref{QGK})-(\ref{QGK1}) we calculate term in electric current proportional to the external electric field strength $\mathcal{F}^{\mu\nu}$:
\begin{widetext}
\begin{eqnarray}
{J}^i
&=&    -\frac{1}{4}
\frac{1}{|\mathfrak{D}|} \int \frac{dP^0}{2\pi} \int_{\mathscr{M}} \frac{d^{2}\vec{P}}{(2\pi)^2}\, \sum_{x\in \mathfrak{D}}
\tr\Bigl(\hat{G} \star \partial_{\pi^{\mu}}\hat{Q}  \star\hat{G} \star \partial_{\pi^{\nu}}\hat{Q} \star \hat{G}  \partial_{\pi_{i}}\hat{Q} \Bigr)^{<}\mathcal{F}^{\mu\nu}\nonumber\\
&&
-\frac{1}{4}
\frac{1}{|\mathfrak{D}|} \int \frac{dP^0}{2\pi} \int_{\mathscr{M}} \frac{d^{2}\vec{P}}{(2\pi)^2}\, \sum_{x\in \mathfrak{D}}  \tr\Bigl(\partial_{\pi_{i}}\hat{Q}  \hat{G} \star \partial_{\pi^{\mu}}\hat{Q}  \star\hat{G} \star \partial_{\pi^{\nu}}\hat{Q} \star \hat{G}  \Bigr)^{<}\mathcal{F}^{\mu\nu}.
\end{eqnarray}
\end{widetext}

Then
$$
{J}^i = \sigma^{ij}  \mathcal{F}_{0j} ,
$$
where the conductivity tensor $\sigma^{ij}$ may be given as follows:
\begin{widetext}
\begin{equation}
\sigma^{ij} =  {\frac{1}{4}} \frac{1}{|\mathfrak{D}|} \int \frac{dP^0}{2\pi} \int_{\mathscr{M}} \frac{d^{2}\vec{P}}{(2\pi)^2}\, \sum_{x\in \mathfrak{D}}  \tr\left(\partial_{\pi_{i}}\hat{Q}_W  \left[\hat{G}_W \star \partial_{\rv{\pi_{[0}}}\hat{Q}_W  \star \partial_{\rv{\pi_{j]}}}\hat{G}_W  \right]\right)^< +{\rm c.c.}\label{MAIN}
\end{equation}
\end{widetext}
Here we restore index $W$ for the Weyl symbols. The anti - symmetrization is denoted by $(...)_{[0} (...)_{ j]} =(...)_{0} (...)_{ j} -(...)_{j} (...)_{ 0}  $.
The asymmetric (Hall) part of conductivity is   $\sigma^{ij}_H = (\sigma^{ij}-\sigma^{ji})/2$.

\subsection{Equilibrium limit of Hall conductivity}
\label{SectEquilibrium}

According to the second trace identity the star may be inserted between the two Weyl symbols standing under the trace. We obtain
\begin{widetext}
\be
\bar \sigma^{ij}
=  \rv{-}{\frac{1}{4}}
\frac{1}{|\mathfrak{D}|} \int \frac{dP^0}{2\pi} \int_{\mathscr{M}} \frac{d^{2}\vec{P}}{(2\pi)^2}\, \sum_{x\in \mathfrak{D}}
\tr\left(\partial_{\pi_{i}}\hat{Q}_W \star \hat{G}_W \star \partial_{\pi_{[0}}\hat{Q}_W  \star \hat{G}_W\star \partial_{\pi_{j]}}\hat{Q}_W\star\hat{G}_W\right)^< +{\rm c.c.}
\ee
\end{widetext}
We assume that $\hat Q$ does not depend on time. For the thermal distribution we represent integral as a sum over Matsubara frequencies. By $\Pi$ we denote  Euclidean $2+1$ - momentum, i.e. $\Pi^{3} = \omega$ is Matsubara frequency,  $\Pi^i = \pi^i$ for $i=1,2$.  $\partial_{\pi^0} = -\ii \partial_{\Pi^{3}}$. We substite $i\omega$ instead of $\pi^0$ and obtain Matsubara Green function $G^M$.

The system with the one - particle Hamiltonian $\hat{H}$ gives the real time Green function
$$
G(x_1,x_2,\omega) \equiv \langle x_1|(\omega - \hat{H})^{-1}|x_2\rangle
$$
It produces Advanced, Retarded or time ordered Green function when the integration contour in plane of complex $\omega$ is shifted in a specific way. Feynman propagator is
\begin{equation}
G^\rT ( x, x^{\prime}, \omega)
= \lim\limits_{\eta\rightarrow 0} G(x, x^{\prime}, \omega+\ii\eta \, {\rm sign} \,\omega).
\end{equation}
The retarded Green's  function is
\begin{equation}
G^\rR ( x, x^{\prime}, \omega) = \lim\limits_{\eta\rightarrow 0} G(x, x^{\prime}, \omega+\ii\eta),
\end{equation}
The advanced Green's function is
\begin{equation}
G^\rA ( x, x^{\prime}, \omega) = \lim\limits_{\eta\rightarrow 0} G(x, x^{\prime}, \omega-\ii\eta).
\end{equation}
The Matsubara Green's function $G^\rM $ is obtained as
\be
G^\rM ( x, x^{\prime},\omega_n) = G( x, x^{\prime},\ii\omega_n),
\label{Mats}
\ee
The latter may be written in terms of imaginary time $\tau$:
\begin{equation}
G^\rM ( x, x^{\prime},\tau)
= \frac{1}{\beta}\sum\limits_{n=-\infty}^{\infty}e^{-\ii\omega_n \tau}G( x, x^{\prime},\ii\omega_n).
\end{equation}
 $\omega_n = (2n+1)\pi/{\beta}$ is the Matsubara frequency, $\beta= 1/T$.

The conductivity averaged over the system area is
$$
\bar{\sigma}^{ij}  = \frac{{\cal N}}{2\pi}\epsilon^{ij},
$$
where
\begin{widetext}
\begin{eqnarray}
{\cal N} &=&2\pi  \, T \, \frac{1}{3! \, }\epsilon^{\mu\nu\rho} \frac{1}{|\mathfrak{D}|}  \int_{\mathscr{M}} \frac{d^{2}\vec{P}}{(2\pi)^2}\, \sum_{x\in \mathfrak{D}}  \sum_{\omega_n = 2\pi T(n+1/2)}
\tr\left(\partial_{\rv{\Pi^{\mu}}}\hat{Q}_W ^\rM  \star\hat{G}_W ^\rM \star \partial_{\rv{\Pi^{\nu}}}\hat{Q}_W ^\rM  \star\hat{G}_W ^\rM \star \partial_{\rv{\Pi^{\rho}}}\hat{Q}_W ^\rM \star \hat{G}_W ^\rM \right).
\end{eqnarray}
\end{widetext}
Here  $\epsilon^{ij}$ and $\epsilon^{\mu\nu\rho}$  are antisymmetric tensors. $Q_W^M$ is inverse to Matsubara Green function $G_W^M$:
$$
Q_W^M \star G_W^M = 1_W  = 1
$$

The sum over Matsubara frequencies is reduced to an integral for small temperatures
\begin{widetext}
\begin{eqnarray}
{\cal N} &=& \frac{1}{3!\,}\epsilon^{\mu\nu\rho} \frac{1}{|\mathfrak{D}|} \int {d\Pi^3}\int_{\mathscr{M}} \frac{d^{2}\vec{\Pi}}{(2\pi)^2}\, \sum_{x\in \mathfrak{D}}
\tr\left(\partial_{\rv{\Pi^{\mu}}}\hat{Q}_W ^\rM  \star\hat{G}_W ^\rM \star \partial_{\rv{\Pi^{\nu}}}\hat{Q}_W ^\rM  \star\hat{G}_W ^\rM \star \partial_{\rv{\Pi^{\rho}}}\hat{Q}_W ^\rM \star \hat{G}_W ^\rM \right).
\label{NEQ}
\end{eqnarray}
\end{widetext}
The sum over $x$ is important for the topological invariance of this quantity.

\section{Conclusions}
\label{SectConcl}

In the present paper we extend the previously proposed construction of precise lattice Wigner - Weyl calculus \cite{FZ2020} to the models defined on the honeycomb lattices. (Recall that in \cite{FZ2020} the rectangular lattices were considered.) Models with artificial honeycomb lattices realize the Hofstadter butterfly, and the quantum Hall effect with effectively large magnetic flux through the lattice cell \cite{polini2013artificial}. For such systems the approximate lattice Wigner - Weyl calculus of \cite{ZZ2019} cannot be used because magnetic flux through the lattice cell appears to be of the order of the quantum of magnetic flux. Then the present construction is inevitable if we are going to represent the QHE conductivity through the topological invariant composed of the Green functions. We derive the corresponding expression. It is manifestly topological invariant, which demonstrates that the QHE conductivity is robust to the smooth modifications of the system.

We started our consideration from the construction of the $\cal B$ symbol that realized the original ideas of F.Buot\footnote{F. Buot's original construction contains a few technical flaws. However, the very notion of such a construction that he offered seems to us to be so significant that we feel it is appropriate to give his name to our corrected construction.}. The $\cal B$ (or Buot) symbol obeys the basic properties of the continuous Wigner - Weyl calculus. However, we observe that the $\cal B$ symbol of unity operator is fast oscillating function of coordinates. As a result the Buot symbol of the Green function does not depend smoothly on coordinates, and the derivative expansion cannot be used in the expression for the electric current. In order to improve the situation we propose the more involved construction with the new symbol of operator, which is called here W symbol or Weyl symbol. It obeys precisely the same properties as Buot symbol, but contrary to the latter, the Weyl symbol of unity operator depends smoothly on coordinates. This allows us to apply derivative expansion to the corresponding expression for electric current. This expansion leads us finally to the expression for the Hall conductivity
\begin{equation}
\sigma_H = \frac{e^2}{h} {\cal N}\label{sigmafin}
\end{equation}
where $\cal N$ is given by Eq. (\ref{NEQ0}).

Thus we have similar constructions of lattice Wigner - Weyl calculus for the rectangular and honeycomb lattices. It would be instructive to extend these constructions to the lattices of arbitrary form. Besides, it would be important to consider effects of interactions. We expect that the latter, being taken into account perturbatively, cannot change the form of Eq. (\ref{NEQ0}), in which the interacting Green function should replace the bare one. The important challenge here is to understand the topological nature of fractional QHE. We suppose that the latter is completely non - perturbative phenomenon. It is not clear at the present moment how the topological expression of Eqs. (\ref{sigmafin}) and (\ref{NEQ0}) is replaced by $e^2/h$ times fractional number. The consideration of these issues, however, remain out of the scope of the present paper.



\bibliographystyle{unsrt}
\bibliography{cross-ref,wigner3,bib_keldysh,wigner4,buotbibl}

\end{document}